\documentclass[twocolumn,final,aps,prr,showpacs,superscriptaddress,amsmath,amssymb,amsfonts,floatfix]{revtex4-2}
\usepackage[utf8]{inputenc}
\usepackage[T1]{fontenc}        
\usepackage{lmodern}            
\usepackage{graphicx}
\usepackage{multirow}
\usepackage{epsfig}
\usepackage{url}
\usepackage{color}

\usepackage[colorlinks,breaklinks,bookmarks=true,citecolor=blue,linkcolor=blue,urlcolor=blue]{hyperref}
\usepackage{color}
\usepackage{tabularx}
\usepackage{sidecap}
\usepackage{float}

\usepackage{orcidlink}

\usepackage{amsmath}            
\usepackage{amstext}            
\usepackage{amssymb}            
\usepackage{mathtools}          
\usepackage{scalerel,stackengine} 
\usepackage{ulem}
\usepackage{xcolor}

\begin{document}

\title{{Closing in on possible scenarios for infinite-layer nickelates:
comparison of dynamical mean-field theory with
angular-resolved photoemission spectroscopy}}

\author{Liang Si\orcidlink{0000-0003-4709-6882}}
\thanks{These authors contributed equally.}
\affiliation{School of Physics, Northwest University, Xi'an 710127, China}
\affiliation{Institute of Solid State Physics, TU Wien, 1040 Vienna, Austria}

\author{Eric Jacob\orcidlink{0009-0005-0940-978X}}
\thanks{These authors contributed equally.}
\affiliation{Institute of Solid State Physics, TU Wien, 1040 Vienna, Austria}

 \author{Wenfeng Wu\orcidlink{0000-0002-6575-5813}}
\thanks{These authors contributed equally.}
\affiliation{Key Laboratory of Materials Physics, Institute of Solid State Physics, HFIPS, Chinese Academy of Sciences, Hefei 230031, China}
\affiliation{Science Island Branch of Graduate School, University of Science and Technology of China, Hefei 230026, China}

 \author{Andreas Hausoel}
\thanks{These authors contributed equally.}
\affiliation{Institute for Theoretical Solid State Physics, Leibniz Institute for Solid State and Materials Research Dresden, Helmholtzstr.\ 20, 01069 Dresden, Germany}

\author{Juraj Krsnik\orcidlink{0000-0002-4357-2629}}
\thanks{These authors contributed equally.}
\affiliation{Institute of Solid State Physics, TU Wien, 1040 Vienna, Austria}

\author{Paul Worm\,\orcidlink{0000-0003-2575-5058}}
\affiliation{Institute of Solid State Physics, TU Wien, 1040 Vienna, Austria}

\author{Simone Di Cataldo\,
\orcidlink{0000-0002-8902-0125}}
\affiliation{Dipartimento di Fisica, Sapienza Universit\`a di Roma, Piazzale Aldo Moro 5, 00187 Roma, Italy}
\affiliation{Institute of Solid State Physics, TU Wien, 1040 Vienna, Austria}

\author{Zhi Zeng}
\affiliation{Key Laboratory of Materials Physics, Institute of Solid State Physics, HFIPS, Chinese Academy of Sciences, Hefei 230031, China}
\affiliation{Science Island Branch of Graduate School, University of Science and Technology of China, Hefei 230026, China}

\author{Oleg Janson} 
\affiliation{Institute for Theoretical Solid State Physics, Leibniz Institute for Solid State and Materials Research Dresden, Helmholtzstr.\ 20, 01069 Dresden, Germany}
 
\author{Karsten Held\orcidlink{0000-0001-5984-8549}}
\email[]{Corresponding authors: siliang@nwu.edu.cn; held@ifp.tuwien.ac.at}
\affiliation{Institute of Solid State Physics, TU Wien, 1040 Vienna, Austria}

\date{\today}

\begin{abstract}
Conflicting theoretical scenarios for infinite-layer nickelate superconductors have been hotly debated, particularly regarding whether {only} a single Ni-3$d_{x^2-y^2}$ band is relevant at low energies besides electron pockets or whether multi-orbital physics including  Ni-3$d_{z^2}$ is instead essential. The first scenario has emerged from density-functional theory plus dynamical mean-field theory (DFT+DMFT) calculations. 
Comparing the previous DFT+DMFT spectra to recent angular-resolved photoemission spectroscopy (ARPES) experiments, we find excellent agreement for both the Fermi surface and the strongly renormalized quasi-particle bands, supporting the first scenario.
Our key findings further suggest that the "waterfalls" observed in ARPES might emerge from the quasi-particle--to--Hubbard-band crossover, and that additional spectral weight close to the $A$-pocket {likely} originates from the Ni-3$d_{xy}$ orbital.
\end{abstract}

\maketitle

\section{Introduction}
Superconductivity in infinite-layer nickelates
has been conjectured \cite{Anisimov1999} due to {their}
close analogy of their electronic structure to cuprates.
Likewise, proposals to engineer heterostructures of perovskite nickelates
aimed at creating an analog of cuprates ~\cite{Chaloupka2008,PhysRevLett.103.016401,Hansmann2010b,PhysRevLett.106.027001}.
Nonetheless, when  superconductivity in
Sr$_x$$R_{1-x}$NiO$_2$
($R$=Nd, La, or Pr) was finally
found \cite{li2019superconductivity,Li2020,zeng2020,Osada2020,zeng2022superconductivity,pan2021,Osada2021,Wang2022}, striking differences
in its electronic structure became apparent in 
density-functional theory (DFT)
calculations \cite{Botana2019,Hirofumi2019,jiang2019electronic,Motoaki2019,hu2019twoband,Wu2019,Nomura2019,Zhang2019,Jiang2019,Werner2019,Si2020,Nomura2022,Kitatani2023b}
and also from its
negative Hall conductivity \cite{li2019superconductivity,zeng2020,Osada2020,zeng2022superconductivity,lee2023linear}, which contrasts with the positive Hall conductivity observed in cuprate superconductors \cite{PhysRevLett.92.197001,balakirev2003signature,badoux2016change}.

At the DFT level, a (predominately) Ni-3$d_{x^2-y^2}$ band crosses the Fermi energy ($E_F$), similar to cuprates, but in addition, electron pockets around the $A$ and $\Gamma$ momenta are present.
Quite importantly, the $A$- and $\Gamma$-pockets, that originate from the $R$ 5$d$-orbitals,
lead to a self-(hole-)doping of the Ni 3$d_{x^2-y^2}$ bands, even for $x$=0. This self-doping can explain the absence of long-range (anti-)ferromagnetism; and the negative Hall conductivity \cite{li2019superconductivity,zeng2020,Osada2020,zeng2022superconductivity,lee2023linear} naturally arises from the electron pockets overcoming the positive Hall contribution of the hole-like Ni-3$d_{x^2-y^2}$ band.

With the partially occupied 3$d$-orbitals, nickelates are firmly in the category of strongly correlated electron systems, where {interactions between electrons} can significantly alter the DFT picture.
{A proper account of} electronic
correlations is however challenging; and it is perhaps not surprising that different scenarios
for the relevant low-energy orbitals have been proposed.
Standard, state-of-the-art DFT+dynamical mean-field theory (DMFT)~\cite{Metzner1989,Georges1992a,Georges1996,held2007electronic} upholds the DFT picture, indicating that only the  Ni-3$d_{x^2-y^2}$, the $A$-pocket, and --for some $R$ and $x$-- the $\Gamma$ pocket cross $E_F$ for the nickelate parent compounds and in the superconducting doping range \cite{Kitatani2020,Karp2020,LaBollita2022b,Pascut2023}. 
Here, electronic correlations strongly renormalize the Ni-3$d_{x^2-y^2}$ band and {shift} the
$\Gamma$-pocket {to higher energies}, so that 
{it entirely disappears} 
for 
LaNiO$_2$ (but not for NdNiO$_2$ \cite{Kitatani2020}). The {alike} superconductive properties of the La and Nd compounds suggest that the $\Gamma$-pocket is not crucial for superconductivity.

Furthermore, since the $A$-pocket does not hybridize with the Ni-3$d_{x^2-y^2}$ band, the arguably simplest model for nickelate superconductivity is (1)
a one-band Hubbard model describing the Ni-3$d_{x^2-y^2}$ band 
plus decoupled (and weakly correlated) pocket(s)~\cite{Kitatani2020,Held2022}. These pocket(s) are crucial for the self-doping of the Ni-3$d_{x^2-y^2}$ orbital but, regarding the effects of correlations and, presumably, superconductivity, act as passive bystanders. 
Naturally, the pockets are relevant for the (Hall) conductivity. Scenario (1) successfully predicted, using the dynamical vertex approximation (D$\Gamma$A) \cite{Toschi2007,RMPVertex}, the superconducting phase diagram of infinite-layer nickelates
\cite{Kitatani2020}, later confirmed experimentally \cite{Lee2023} \footnote{For a one-to-one comparison, see \cite{Worm2024}}.
Additionally, the spin excitations in scenario (1), which act as the pairing mechanism for superconductivity, conform well with resonant inelastic X-ray spectroscopy (RIXS) \cite{Lu2021,Worm2024}.

Other scenarios \cite{Adhikary2020,Wang2020t,WangZ2020,Kreisel2022} emphasize (2) the role of holes in the  Ni-3$d_{z^2}$ orbital and associated multi-orbital physics arising from the interaction between Ni-3$d_{z^2}$ and Ni-3$d_{x^2-y^2}$. These holes
{emerge
because} Ni-3$d_{z^2}$ hybridizes strongly with the $R$-5$d_{z^2}$ orbitals
around 
the $\Gamma$-pocket.
Scenario (3) proposes an even more pronounced role of the Ni-3$d_{z^2}$ orbital, suggesting it forms an {\it additional}
Ni-3$d_{z^2}$ Fermi surface.
Variants of scenario (3) appear in self-interaction corrected (sic) DFT+DMFT
\cite{Lechermann2019,Lechermann2020,Lechermann2021}, $GW$+DMFT \cite{Petocchi2020}, and anti-ferromagnetically ordered DFT \cite{Wan2021,MiYoung2020}.
Such multi-orbital physics and Fermi surfaces are also found in standard DFT+DMFT, but only in the over-doped regime \cite{Kitatani2020}. 
The difference between the DFT+DMFT and sicDMFT+DMFT results, with $GW$+DMFT somewhat in the middle, is the crystal field splitting between  Ni-3$d_{z^2}$  and   Ni-3$d_{x^2-y^2}$ orbital, a smaller splitting between these two orbitas leads to multi-Ni-orbital physics already at smaller hole dopings.

Thanks to recent angular-resolved photoemission spectroscopy (ARPES) experiments for infinite-layer nickelates by Sun {\it et al.}~\cite{Sun2024} and Ding {\it et al.}~\cite{Ding2024}, we are eventually in the situation to discriminate between the different theory scenarios. These ARPES experiments employ molecular beam epitaxy (MBE) and {\it in-situ} topotactical reduction, which avoids exposing the films to air \footnote{as in earlier photoemission spectroscopy (PES) experiments \cite{Chen2022}}.
In the present paper, we compare the original DFT+DMFT results \cite{Kitatani2020} to ARPES, {showing an almost perfect agreement}.
We also identify the likely origin of the experimentally observed waterfalls and the additional spectral weight around the $A$-momentum.

\section{Method}
Here, we use the self-energy from previous DFT+DMFT calculations \cite{Kitatani2020} to obtain the energy- and momentum-resolved spectral function $A$($\mathbf{k},\omega$)
in the same frequency and momentum range as in ARPES. Key points of these calculations include: \textsc{wien2k} \cite{blaha2001wien2k} with GGA-PBEsol \cite{PhysRevLett.100.136406} exchange correlation potential for DFT at DFT-relaxed lattice parameters  $a$=$b$=3.89\,\AA, $c$=3.35\,\AA;
projection onto {all} Ni-3$d$ and La-5$d$ orbitals using \textsc{wien2wannier} \cite{Kunes2010a} and {constructing maximally localized Wannier functions \cite{RevModPhys.84.1419} with \textsc{wannier90} \cite{mostofi2008wannier90,Pizzi2020}}.
{The Coulomb repulsion $U'$=3.10\,eV (2.00\,eV) and Hund's exchange   $J$=0.65\,eV (0.25\,eV) for Ni-3$d$ (La-5$d$) orbitals have been calculated  \cite{Si2020}  by constrained random phase approximation (cRPA) \cite{PhysRevB.77.085122}.} Here, $U'$ is the local inter-orbital Coulomb interaction, the intra-orbital interaction is given by $U=U'+2J$.
Using these orbitals and interactions, along with the standard double counting correction \cite{Anisimov1991},  DMFT~\cite{Georges1996,Kotliar2006,held2007electronic} calculations at room temperature (300\,K) were performed, employing continuous-time quantum Monte Carlo (QMC) in the hybridization expansion \cite{Gull2011a} as implemented in \textsc{W2dynamics} \cite{w2dynamics2018}. The maximum entropy \cite{PhysRevB.44.6011} with the \textsc{ana\_cont} code \cite{Kaufmann2021} was used for the analytic continuation. Sr doping of 20\% was taken into account in a rigid-band model in DFT{;} 
{within DMFT the changed filling changes the strength of the electronic correlations and alters the spectrum beyond a mere shift.}

We here build upon the previously calculated DMFT self-energy  (except for Fig.~\ref{Fig:Waterfalls}
that has improved statistics)   to ensure that there has been no parameter adjustment,
{This guarantees  that the theoretical results are genuine predictions.}
From the DFT-derived low-energy Hamiltonian $H{(\mathbf k)}$, the DMFT self-energy $\Sigma(\omega)$ and chemical potential $\mu$ was calculated {\it before} experiments \cite{Sun2024,Ding2024}. In the present paper, the spectra are simply obtained as: 
\begin{equation}
    \label{eq:akw}
    A({\mathbf k}, \omega)= -\frac{1}{\pi}{\operatorname{Tr}}\left({\rm Im} \left[\omega+\mu- H{(\mathbf k)} - \Sigma(\omega)\right]^{-1}\right)\;.
\end{equation}
Here, all quantities are to be understood as matrices in the {basis of} 5 Ni-3$d$ and 5 La-5$d$ orbitals, with $\omega$ and $\mu$ being scalars multiplied by the unit matrix.
{Energies (momenta) are measured {in units} of eV (inverse lattice constant for the respective direction).}

Let us mention that, at large energies (such as those for the waterfalls below), we do not expect a major difference
between DMFT and D$\Gamma$A or other methods including non-local correlations because spin fluctuations  have a relatively low energy scale of only $\sim 100\,$meV. For the same reason, we do not expect a difference in the overall quasiparticle renormalization. However, at low energies, non-local correlations can lead to some deviations of the overall renormaliztation and $\mathbf k$-dependence of this quasiparticle renormalization, they can somewhat reshape (flatten) the Fermi surface, and at Sr-dopings considerably smaller than 20\%, they can also also lead to the opening of a pseudogap \cite{Kitatani2020}. All of these effects are however in the details, most likely comparable to the error of the ARPES experiment
that is slightly different from Brillouin zone quadrant to Brillouin zone  quadrant and cannot resolve the superconducting gap.

\section{Results}
\subsection{Fermi surface}
\begin{figure}[t]
\includegraphics[width=8.5cm]{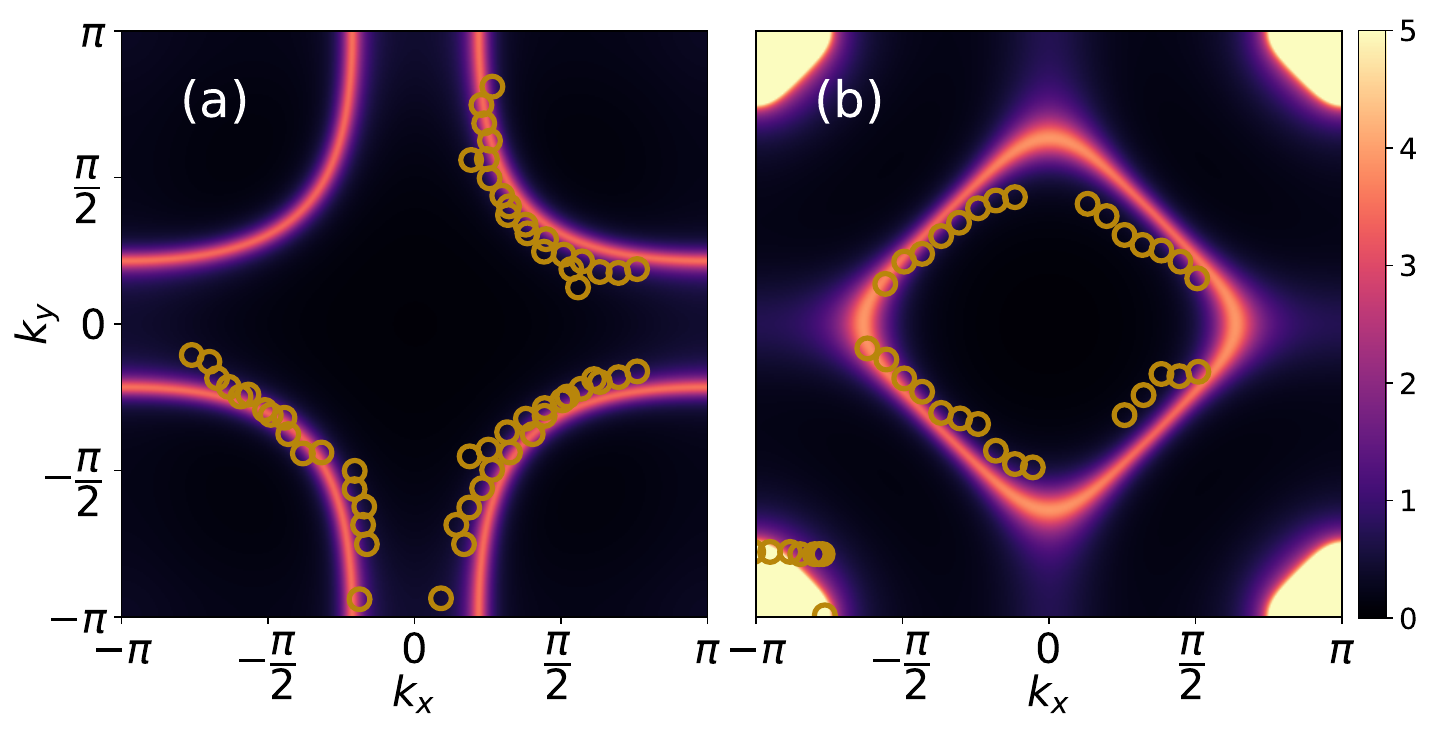}
\caption{DFT+DMFT spectral function for Sr$_{0.2}$La$_{0.8}$NiO$_2$ at $E_F$ (false colors in units of eV$^{-1}$) compared to ARPES  from \cite{Sun2024} (circles) for (a) $k_z=0$ and (b) $k_z=\pi$. The Ni-3$d_{x^2-y^2}$ Fermi surface changes from (a) hole-like to (b) electron-like plus an additional (filled) pocket around the $A=(\pi,\pi,\pi)$ momentum. \label{Fig:FS}}
\end{figure}

Fig.~\ref{Fig:FS} compares the DFT+DMFT spectral function $A({\mathbf k},\omega)$ at $E_F$
to ARPES data (circles) for two different momenta $k_z$.
For (a) $k_z=0$ both theory and experiment show a single hole-like Fermi surface originating from the Ni-3$d_{x^2-y^2}$ band, very similar to cuprates. At (b) $k_z=\pi$, this Ni-3$d_{x^2-y^2}$ Fermi surface becomes electron-like, and an additional pocket emerges at the $A=(\pi,\pi,\pi)$  momentum.

\begin{figure}[t]
\includegraphics[width=8.5cm]{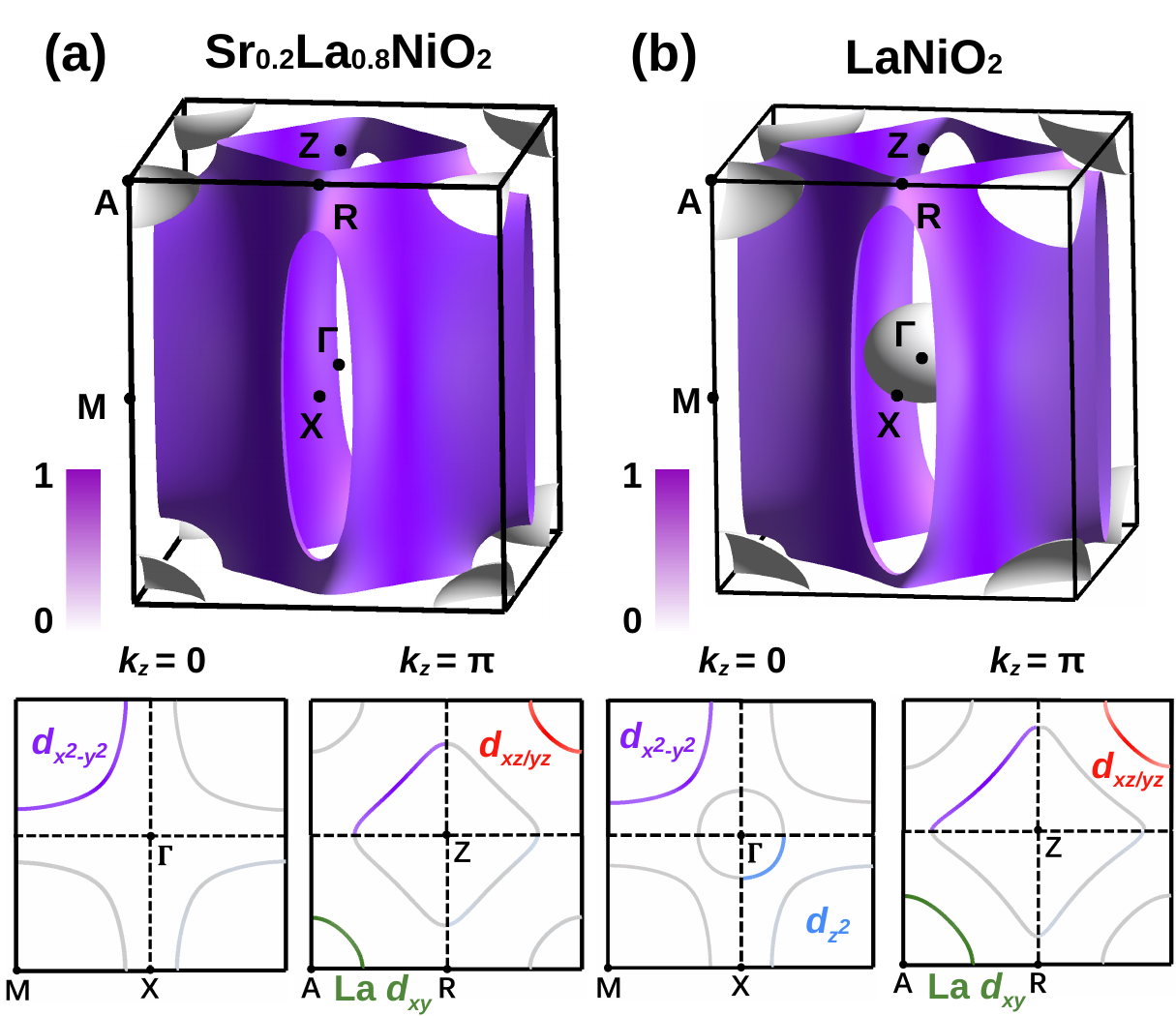}
\caption{DFT Fermi surface for (a) Sr$_{0.2}$La$_{0.8}$NiO$_2$ and (b) LaNiO$_2$. Electronic correlations remove the $\Gamma$-pocket found for LaNiO$_2$ in DFT.
In the top panels, the  colorbar of the 3D DFT Fermi surface indicates the contribution of the Ni-3$d_{x^2-y^2}$ orbital. In the 2D Fermi surface for $k_z=0$ and $k_z=\pi$ of the bottom panels,  the purple, red, green and blue  color indicates, in one  quadrant each, the contribution from the Ni-$d_{x^2-y^2}$, Ni-$d_{yz}$/$d_{xz}$, La-$d_{xy}$ and Ni-$d_{z^2}$ orbital, respectively.}
\label{Fig:FSDFT}
\end{figure}

{{Also} in experiments}, the $A$-pocket appears more as a filled patch than a {hollow} circle~\footnote{See Fig.~1 of \cite{Sun2024} and Fig.~1 of \cite{Ding2024}}. Therefore, for the A pocket, the second derivative of the moment distribution curve, rather than its  maximum (as for the other parts of the Fermi surface), was used to determine the circles in Fig.~\ref{Fig:FS} \cite{Ding2024}. The possible nature of this filled $A$-pocket will be discussed below.
Except for this last aspect, DFT also agrees with the DFT+DMFT Fermi surface for Sr$_{0.2}$La$_{0.8}$NiO$_2$, see Fig.~\ref{Fig:FSDFT}(a). In particular, both DFT and DFT+DMFT agree with experimental observations that there is no additional Fermi surface from the Ni-3$d_{z^2}$ band. This {is at odds with} scenarios of type (3) that entail an additional Ni-3$d_{z^2}$ Fermi surface~\footnote{See e.g.~Fig.~11 of \cite{Lechermann2020} and Fig.~4(f) of \cite{Petocchi2020}}.

For the undoped parent compound, DFT {reveals, besides} the Ni-3$d_{x^2-y^2}$ Fermi surface, {an additional} $\Gamma$-pocket, as illustrated in Fig.~\ref{Fig:FSDFT}(b). In contrast, scenario (3) {features} a fragmented Fermi surface around $\Gamma$ \footnote{See Fig.~3(b) of \cite{Lechermann2020} and Fig.~4(a) of \cite{Petocchi2020}}, but lacks a cuprate-like hole Fermi surface at $k_z=0$.
ARPES experiments for LaNiO$_2$ \cite{Ding2024} and DFT+DMFT \cite{Kitatani2020} 
instead show exactly this cuprate-like Ni-3$d_{x^2-y^2}$ Fermi surface and no feature around $\Gamma$, qualitatively similar to Sr$_{0.2}$La$_{0.8}$NiO$_2$ in Fig.~\ref{Fig:FS}(a).

The absence of the $\Gamma$-pocket is readily explained in scenario (1) {as a correlation-driven} shift of the $\Gamma$ pocket (but not the $A$-pocket) above $E_F$.
The absence of the $\Gamma$-pocket also speaks against scenarios of type (2). In this scenario, the Ni-3$d_{z^2}$ orbitals are not fully filled because they strongly hybridize with the La band around $\Gamma$. 
Due to this depopulation, scenarios of type (2) conjecture the interaction between Ni-3$d_{z^2}$ and Ni-3$d_{x^2-y^2}$ electrons to be a key factor. However, the experiment shows no $\Gamma$ pocket, hence there are no Fermi surface sheets related to the Ni-3$d_{z^2}$ orbital. {An} active role of this orbital 
is {thus} difficult to conceive. 

Quite generally, any fully filled and fully empty band(s) will not contribute to the low-energy physics. Hence, without a Ni-3$d_{z^2}$ Fermi surface as in scenario (3) nor with a $\Gamma$-pocket to which the Ni-3$d_{z^2}$ contributes by hybridization with La-$d_{z^2}$ as in scenario (2), the Ni-3$d_{z^2}$ orbital cannot be relevant for physics at the low energies.

Besides the Ni-3$d_{x^2-y^2}$ Fermi surface, there is also the $A$-pocket in ARPES and DFT+DMFT. However, both do not hybridize. Both could couple by Coulomb interaction, but this a weaker effects also since the admixing of other Ni orbitals to the $A$-pocket is smaller than for the $\Gamma$-pocket. This is the reason why it has not been proposed in any multi-orbital scenario.
As a matter of course, even a decoupled $A$-pocket as in scenarion (1) contributes to the (Hall) resistivity, but as a separate contribution not through actual (coupled) multi-orbital physics.
Altogether, multi-orbital effects--certainly at least those multi-orbital physics that has been proposed--can be ruled out given the ARPES Fermi surfaces, which agree with DFT+DMFT and scenario (1).

\subsection{Band dispersion}
\begin{figure*}[t]
  \begin{minipage}{.66\textwidth}
\includegraphics[width=.99\textwidth,angle=0]{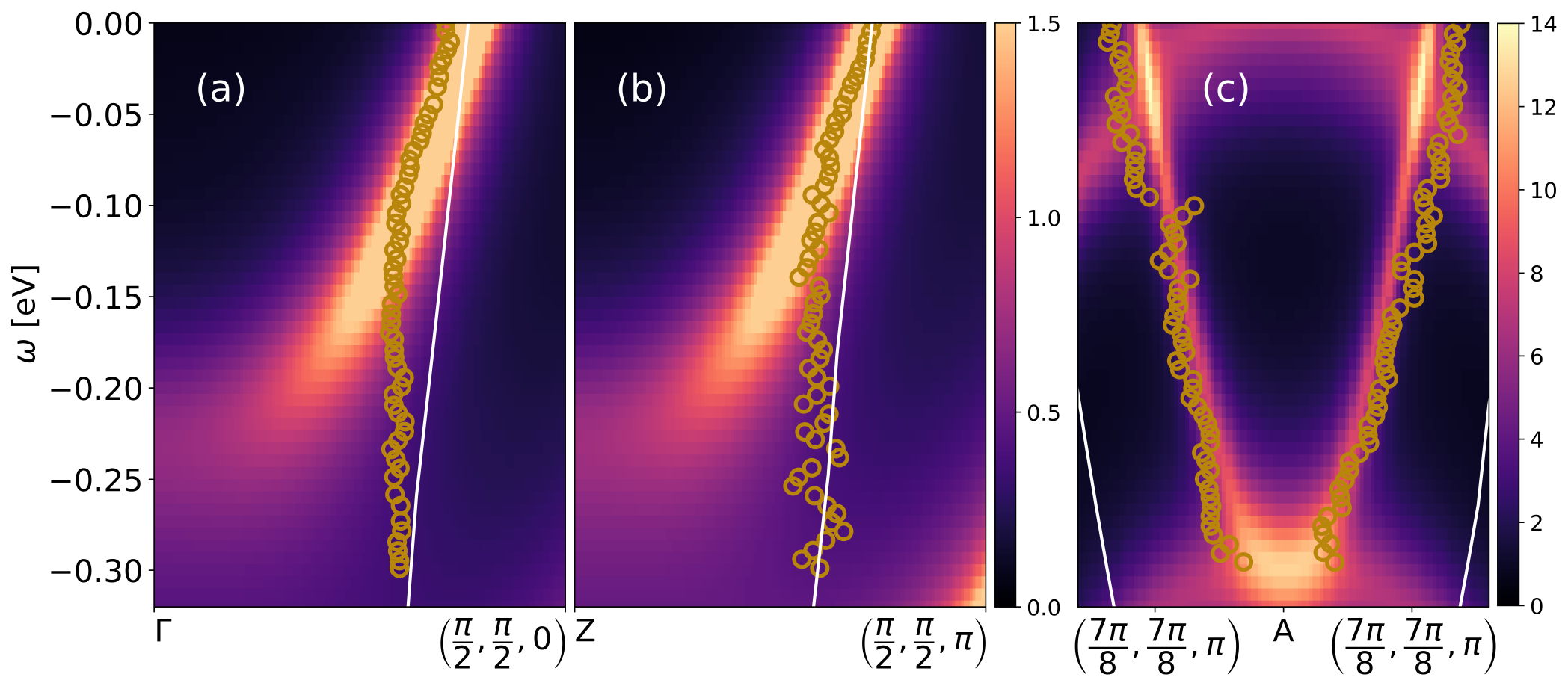}
\end{minipage}\hfill
\begin{minipage}{.33\textwidth}
\caption{DFT+DMFT spectral function $A({\mathbf k},\omega)$ for Sr$_{0.2}$La$_{0.8}$NiO$_2$ (false colors) compared to DFT (white lines) and ARPES from \cite{Sun2024} (golden circles) for three different ${\mathbf k}$ paths. Electronic correlations strongly renormalize the dispersion of the Ni-3$d_{x^2-y^2}$ band (a,b), while the dispersion of the $A$-pocket {in (c)} is hardly renormalized.}
\label{Fig:Dispersion}
\end{minipage}
\end{figure*}

Electronic correlations have a much more pronounced effect on the energy-momentum dispersion than on the Fermi surface. Specifically, they induce a quasi-particle renormalization, which was predicted in DFT+DMFT to be 2.9 for the Ni-3$d_{x^2-y^2}$ band {in}  Sr$_{0.2}$La$_{0.8}$NiO$_2$ \cite{Kitatani2020}. This prediction aligns excellently with the ARPES for energies close to $E_F$, see Fig.~\ref{Fig:Dispersion}(a,b) for $k_z=0$ and $\pi$, respectively.  A striking feature is a waterfall-like dispersion observed in the experiment below $-0.1$\,eV \cite{Sun2024}. This happens at a point where the DFT+DMFT spectrum in  Fig.~\ref{Fig:Dispersion} becomes considerably smeared out{; and}
we will turn to a possible explanation for this waterfall below.

In contrast to the strong quasi-particle renormalization observed in the Ni-3$d_{x^2-y^2}$ band at low energies, the dispersion of the $A$-pocket in  Fig.~\ref{Fig:Dispersion}(c) remains essentially unrenormalized by electronic correlations. {However} it is shifted, which is essential for the excellent agreement between DMFT and ARPES for the $A$-pocket in  Fig.~\ref{Fig:Dispersion}(c). 
{Note that the DFT $A$-pocket is much larger; the white line in  Fig.~\ref{Fig:Dispersion}(c) is mostly outside the displayed ${\mathbf k}$-$\omega$ range.}
Given the striking difference in the strength of correlations between the strong correlations of the Ni-3$d_{x^2-y^2}$ band and very weak correlations of the $A$-pocket, plus the fact that they do not hybridize, it was assumed in \cite{Kitatani2020} that unconventional superconductivity arises from the Ni-3$d_{x^2-y^2}$ band. The $A$-pocket then plays the role of passive bystander as far as correlations and, prospectively, superconductivity are concerned.

\subsection{Filled Fermi surface at A}
Let us now {explore}
why the $A$-pocket appears as a filled circle in both, ARPES \cite{Sun2004,Ding2024} and DFT+DMFT [Fig.~\ref{Fig:FS}(b)], rather than an open circle as predicted in DFT  in Fig.~\ref{Fig:FSDFT} and \cite{Botana2019,Hirofumi2019,jiang2019electronic,Motoaki2019,hu2019twoband,Wu2019,Nomura2019,Zhang2019}.
{At least in DMFT,} this ``filling'' is due to an additional hole-like band touching $E_F$ as shown in Fig.~\ref{Fig:Dispersion}(c).
This band is composed of predominant Ni-3$d_{xy}$ character, resulting in a relatively flat dispersion and short lifetimes, indicative of significant smearing due to the interactions with electrons in the  Ni-3$d_{x^2-y^2}$ orbital. 
That this band touches $E_F$ is a peculiarity of {Sr$_x$La$_{1-x}$NiO$_2$ at $x\gtrsim 0.2$; for Sr$_x$Nd$_{1-x}$NiO$_2$ and smaller $x$ it remains well below $E_F$ \cite{Kitatani2020}. 
Hence, it cannot be essential   for superconductivity. 
}

{
The presence of the filled $A$-pocket in the {\it a priori}  DMFT calculation 
speaks in favor of this interpretation.
Let us however also note the interpretations as an additional band around $A$ suggested in  \cite{Sun2024,DingV1},
albeit \cite{Sun2024} and \cite{DingV1} arrived at opposite slopes for this conjectured, but difficult to discern extra band.
Further note that the exact position of the Ni-3$d_{xy}$ band relative to the Fermi energy is sensitive to various factors such as the already mentioned choice of La or Nd, doping  with Sr or Ca  as well as substrate strain and surface effects. }

\subsection{Waterfalls}
A remarkable feature of the ARPES spectrum is the waterfall-like dispersion observed in the experiment, see circles Fig.~\ref{Fig:Dispersion}(a,b).
This phenomenon has been interpreted in Refs.~\cite{Sun2024,Ding2024} as arising from interactions with a bosonic mode, such as phonons or spin fluctuations.
However, phonon energies in pure nickelates are restricted to below 80\,meV; with only topotactic hydrogen phonon modes reaching up to 190\,meV \cite{Si2022a,PhysRevB.107.165116,DiCataldo2023}.
Similarly, spin fluctuations are restricted to below 200\,meV for undoped nickelates and even lower energies upon Sr-doping \cite{Lu2021,Worm2024}.
While these energies {might} agree with the onset of the waterfall feature in Fig.~\ref{Fig:Dispersion}, the waterfall extends over a much larger energy window, exceeding 300\,meV. This is the end of the experimental energy window of \cite{Sun2024}.
In cuprates it has been observed that waterfalls only end at the order of 1\,eV binding energy.
It is {difficult} to imagine that  bosonic modes at around 100-200\,meV can influence the dispersion at such high binding energies.

\begin{figure}[t]
\includegraphics[width=8.75cm, angle=0]{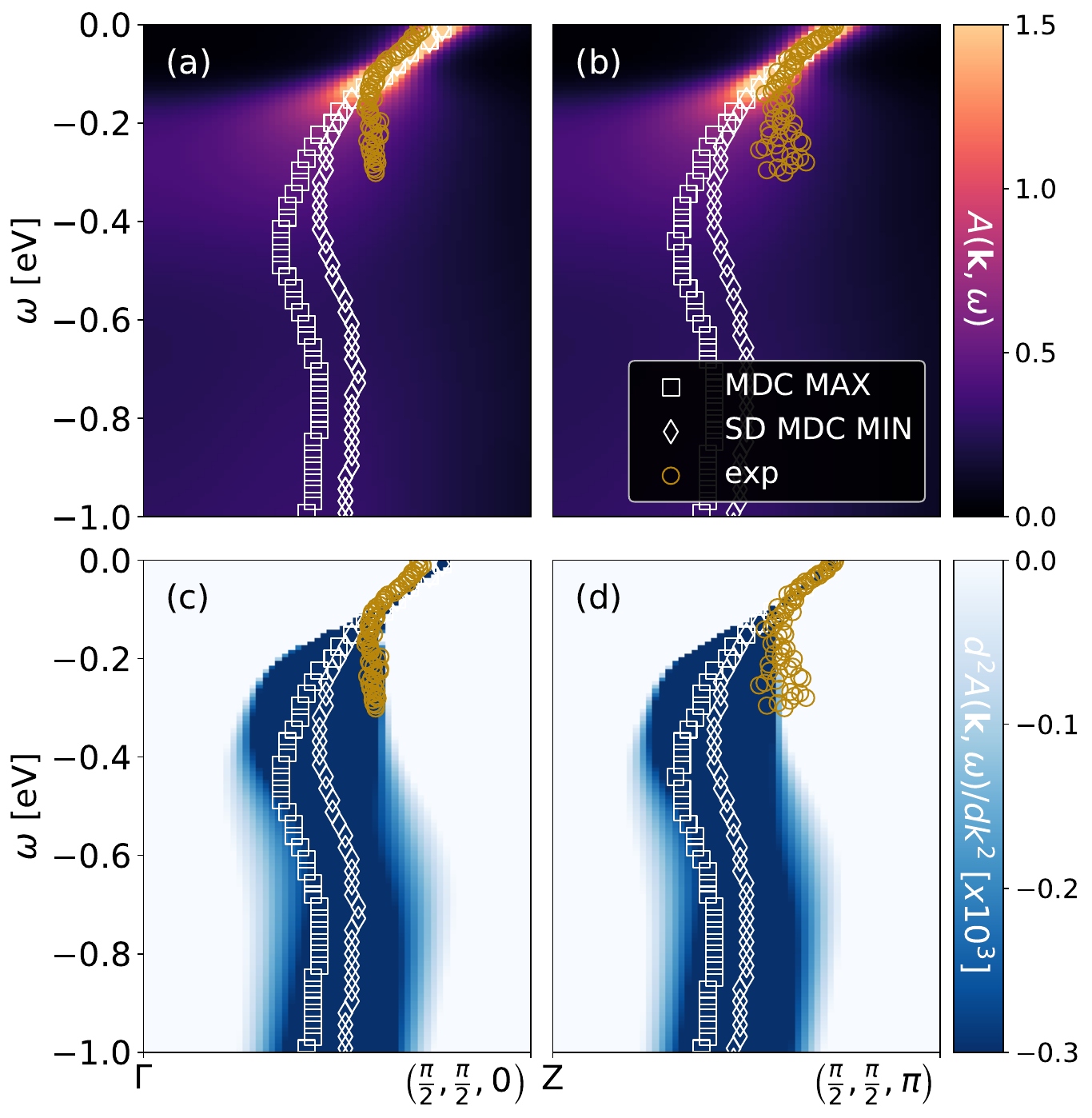}
\caption{(a,b) Waterfalls in the Ni-3$d_{x^2-y^2}$ band of Sr$_{0.2}$La$_{0.8}$NiO$_{2}$ with improved QMC statistics in multi-orbital DFT+DMFT, compared to ARPES from \cite{Sun2024} (golden circles), along two ${\mathbf k}$-paths. 
(c,d) The waterfalls are more clearly visible in the second derivative (SD) of the momentum distribution curve (MDC).}
\label{Fig:Waterfalls}
\end{figure}

Here we arrive at an alternative explanation.
Already in Fig.~\ref{Fig:Dispersion}(a,b), we see that, at the onset of the {experimental} waterfall, the Ni-3$d_{x^2-y^2}$ band fades away{.
Moreover, the waterfall is present in {single-Ni-3$d_{x^2-y^2}$-orbital DMFT calculation  and was reported before}  in DMFT for nickelates under pressure \cite{DiCataldo2023b}}.
{A careful analysis with improved QMC statics of $10^9$ measurements also allows us to resolve a waterfall{-like structure} in Fig.~\ref{Fig:Waterfalls}(a,b).
In particular when taking} the second derivative
$\partial{^2}A({\mathbf k},\omega)/\partial {\mathbf k}^2$, as in experiment, we
see a {waterfall} in multi-orbital DFT+DMFT calcualtion for nickelates presented in Fig.~\ref{Fig:Waterfalls}(c,d). For the Hubbard model, it was demonstrated that these waterfalls are quite robust at intermediate coupling and 20\% hole doping. In some parameter-regimes, they are also visible directly in $A({\mathbf k},\omega)$ not only in its second derivative \cite{Krsnik2024}. 

The waterfalls of Fig.~\ref{Fig:Waterfalls}(c,d) are a local correlation effect already contained in DMFT, connecting the quasiparticle band to the lower Hubbard band. Similar waterfall-like structures have {also} been observed in cluster extensions of DMFT and lattice QMC before \cite{Macridin2007,Moritz2010}, albeit in these numerical calculations it is unclear whether spin fluctuations are at the origin, leading to {ambiguous} interpretations.

The fact that there are waterfalls within our---also otherwise---quite accurate description of infinie-layer nickelates and that these waterfalls induced by  local correlation must be present at intermediate coupling \cite{Krsnik2024}, speaks in favor of our physical meachanism instead of the coupling to some bosonic mode.
The true litmus test will be to see whether, in experiment, they also extend down to 1\,eV binding energy as in cuprates.  This would evidence the local correlation scenario and rule out, on the other hand, the coupling to a boson mode  with energy 100-200\,meV.

\section{Conclusion}
At long last, the controversy regarding single- vs.\ multi-orbital physics in infinite-layer nickelates  nears closure. The shown one-to-one comparison of DFT+DMFT and ARPES, together with the absence of a $\Gamma$-pocket and no Ni-3$d_{z^2}$ Fermi surface, essentially rule out the multi-orbital scenario as the key driver for superconductivity in nickelates. Qualitatively, and to a surprising extent also quantitatively, superconductivity, spin-fluctuations, and ARPES spectra can be described by a correlated Ni-3$d_{x^2-y^2}$ band plus $A$-pocket(s). The $d$-wave superconductivity \footnote{Also experimentally, the evidence towards $d$-wave superconductivity tightens \cite{Harvey2022,Chow2022}, while --at the beginning-- also mixed $d$-and $s$-wave superconductivity has been suggested \cite{Gu2020b}.} in infinite-layer nickelates can be explained theoretically with spin-fluctuations that agree in strength with the experiment. Here, we have seen that  the theoretical electronic structure of these calculations is akin to ARPES as well.
This does not rule out that multi-orbital effects such as the additional feature at the $A$ momentum, hybridization, and super-exchange processes with the bands above and below the Fermi energy, {can} be in charge of minor quantitative corrections or at the origin of prospectively different $T_C$'s for different rare earth elements \footnote{The extent to which the intrinsic $T_C$ differs in experiment beyond likely differences in the quality of the films is not clear at the moment.}. However, the good agreement of DFT+DMFT and ARPES  essentially rules out multi-Ni-band physics as the {{\it main driver}} for superconductivity in nickelates.

Furthermore, we {have a possible explanation for} the additional feature at the $A$ momentum: 
it {might originate} from the Ni-3$d_{xy}$ orbital touching the Fermi energy, a peculiarity of Sr$_x$La${_{1-x}}$NiO$_2$, not observed in DFT+DMFT for the Nd and Pr compounds, which can be checked experimentally.
Finally, we showed that the observed waterfalls naturally arise from local DMFT correlations. They originate from the crossover between the quasi-particle band to the lower Hubbard band. The overall good agreement with the experiment suggests this mechanism is behind the waterfalls in nickelates and, by analogy, likely also in cuprates \cite{PhysRevLett.99.237002,PhysRevLett.98.067004,PhysRevLett.98.147001,PhysRevB.75.174506,PhysRevB.71.094518}.

{\it Acknowledgments.}
We thank Motoharu Kitatani and Leonard Verhoff
for helpful discussions.
We further acknowledge funding by the Austrian Science Funds (FWF) through projects I~5398 {(DFG project ID 465000489)}, SFB Q-M\&S (FWF project ID F86),  and Research Unit
QUAST by the Deutsche Foschungsgemeinschaft (DFG project ID FOR5249; FWF project ID I 5868). L.S.~is thankful for the starting funds from Northwest University. The (previous) DFT+DMFT calculations have been mainly done on the Vienna Scientific Cluster (VSC).

For the purpose of open access, the authors have applied a CC BY public copyright license to any Author Accepted Manuscript version arising from this submission.

 

\begin{thebibliography}{94}%
\makeatletter
\providecommand \@ifxundefined [1]{%
 \@ifx{#1\undefined}
}%
\providecommand \@ifnum [1]{%
 \ifnum #1\expandafter \@firstoftwo
 \else \expandafter \@secondoftwo
 \fi
}%
\providecommand \@ifx [1]{%
 \ifx #1\expandafter \@firstoftwo
 \else \expandafter \@secondoftwo
 \fi
}%
\providecommand \natexlab [1]{#1}%
\providecommand \enquote  [1]{``#1''}%
\providecommand \bibnamefont  [1]{#1}%
\providecommand \bibfnamefont [1]{#1}%
\providecommand \citenamefont [1]{#1}%
\providecommand \href@noop [0]{\@secondoftwo}%
\providecommand \href [0]{\begingroup \@sanitize@url \@href}%
\providecommand \@href[1]{\@@startlink{#1}\@@href}%
\providecommand \@@href[1]{\endgroup#1\@@endlink}%
\providecommand \@sanitize@url [0]{\catcode `\\12\catcode `\$12\catcode `\&12\catcode `\#12\catcode `\^12\catcode `\_12\catcode `\%12\relax}%
\providecommand \@@startlink[1]{}%
\providecommand \@@endlink[0]{}%
\providecommand \url  [0]{\begingroup\@sanitize@url \@url }%
\providecommand \@url [1]{\endgroup\@href {#1}{\urlprefix }}%
\providecommand \urlprefix  [0]{URL }%
\providecommand \Eprint [0]{\href }%
\providecommand \doibase [0]{https://doi.org/}%
\providecommand \selectlanguage [0]{\@gobble}%
\providecommand \bibinfo  [0]{\@secondoftwo}%
\providecommand \bibfield  [0]{\@secondoftwo}%
\providecommand \translation [1]{[#1]}%
\providecommand \BibitemOpen [0]{}%
\providecommand \bibitemStop [0]{}%
\providecommand \bibitemNoStop [0]{.\EOS\space}%
\providecommand \EOS [0]{\spacefactor3000\relax}%
\providecommand \BibitemShut  [1]{\csname bibitem#1\endcsname}%
\let\auto@bib@innerbib\@empty
\bibitem [{\citenamefont {Anisimov}\ \emph {et~al.}(1999)\citenamefont {Anisimov}, \citenamefont {Bukhvalov},\ and\ \citenamefont {Rice}}]{Anisimov1999}%
  \BibitemOpen
  \bibfield  {author} {\bibinfo {author} {\bibfnamefont {V.~I.}\ \bibnamefont {Anisimov}}, \bibinfo {author} {\bibfnamefont {D.}~\bibnamefont {Bukhvalov}},\ and\ \bibinfo {author} {\bibfnamefont {T.~M.}\ \bibnamefont {Rice}},\ }\bibfield  {title} {\bibinfo {title} {{Electronic structure of possible nickelate analogs to the cuprates}},\ }\href {https://doi.org/10.1103/PhysRevB.59.7901} {\bibfield  {journal} {\bibinfo  {journal} {Phys. Rev. B}\ }\textbf {\bibinfo {volume} {59}},\ \bibinfo {pages} {7901} (\bibinfo {year} {1999})}\BibitemShut {NoStop}%
\bibitem [{\citenamefont {Chaloupka}\ and\ \citenamefont {Khaliullin}(2008)}]{Chaloupka2008}%
  \BibitemOpen
  \bibfield  {author} {\bibinfo {author} {\bibfnamefont {J.}~\bibnamefont {Chaloupka}}\ and\ \bibinfo {author} {\bibfnamefont {G.}~\bibnamefont {Khaliullin}},\ }\bibfield  {title} {\bibinfo {title} {Orbital order and possible superconductivity in lanio$_{3}$/lamo$_{3}$ superlattices},\ }\href {https://doi.org/10.1103/PhysRevLett.100.016404} {\bibfield  {journal} {\bibinfo  {journal} {Physical Review Letters}\ }\textbf {\bibinfo {volume} {100}},\ \bibinfo {pages} {16404} (\bibinfo {year} {2008})}\BibitemShut {NoStop}%
\bibitem [{\citenamefont {Hansmann}\ \emph {et~al.}(2009)\citenamefont {Hansmann}, \citenamefont {Yang}, \citenamefont {Toschi}, \citenamefont {Khaliullin}, \citenamefont {Andersen},\ and\ \citenamefont {Held}}]{PhysRevLett.103.016401}%
  \BibitemOpen
  \bibfield  {author} {\bibinfo {author} {\bibfnamefont {P.}~\bibnamefont {Hansmann}}, \bibinfo {author} {\bibfnamefont {X.}~\bibnamefont {Yang}}, \bibinfo {author} {\bibfnamefont {A.}~\bibnamefont {Toschi}}, \bibinfo {author} {\bibfnamefont {G.}~\bibnamefont {Khaliullin}}, \bibinfo {author} {\bibfnamefont {O.~K.}\ \bibnamefont {Andersen}},\ and\ \bibinfo {author} {\bibfnamefont {K.}~\bibnamefont {Held}},\ }\bibfield  {title} {\bibinfo {title} {{Turning a Nickelate {Fermi} Surface into a Cupratelike One through Heterostructuring}},\ }\href {https://doi.org/10.1103/PhysRevLett.103.016401} {\bibfield  {journal} {\bibinfo  {journal} {Phys. Rev. Lett.}\ }\textbf {\bibinfo {volume} {103}},\ \bibinfo {pages} {16401} (\bibinfo {year} {2009})}\BibitemShut {NoStop}%
\bibitem [{\citenamefont {Hansmann}\ \emph {et~al.}(2010)\citenamefont {Hansmann}, \citenamefont {Toschi}, \citenamefont {Yang}, \citenamefont {Andersen},\ and\ \citenamefont {Held}}]{Hansmann2010b}%
  \BibitemOpen
  \bibfield  {author} {\bibinfo {author} {\bibfnamefont {P.}~\bibnamefont {Hansmann}}, \bibinfo {author} {\bibfnamefont {A.}~\bibnamefont {Toschi}}, \bibinfo {author} {\bibfnamefont {X.}~\bibnamefont {Yang}}, \bibinfo {author} {\bibfnamefont {O.~K.}\ \bibnamefont {Andersen}},\ and\ \bibinfo {author} {\bibfnamefont {K.}~\bibnamefont {Held}},\ }\bibfield  {title} {\bibinfo {title} {{Electronic structure of nickelates: From two-dimensional heterostructures to three-dimensional bulk materials}},\ }\href {https://doi.org/10.1103/PhysRevB.82.235123} {\bibfield  {journal} {\bibinfo  {journal} {Phys. Rev. B}\ }\textbf {\bibinfo {volume} {82}},\ \bibinfo {pages} {235123} (\bibinfo {year} {2010})}\BibitemShut {NoStop}%
\bibitem [{\citenamefont {Uchida}\ \emph {et~al.}(2011)\citenamefont {Uchida}, \citenamefont {Ishizaka}, \citenamefont {Hansmann}, \citenamefont {Kaneko}, \citenamefont {Ishida}, \citenamefont {Yang}, \citenamefont {Kumai}, \citenamefont {Toschi}, \citenamefont {Onose}, \citenamefont {Arita}, \citenamefont {Held}, \citenamefont {Andersen}, \citenamefont {Shin},\ and\ \citenamefont {Tokura}}]{PhysRevLett.106.027001}%
  \BibitemOpen
  \bibfield  {author} {\bibinfo {author} {\bibfnamefont {M.}~\bibnamefont {Uchida}}, \bibinfo {author} {\bibfnamefont {K.}~\bibnamefont {Ishizaka}}, \bibinfo {author} {\bibfnamefont {P.}~\bibnamefont {Hansmann}}, \bibinfo {author} {\bibfnamefont {Y.}~\bibnamefont {Kaneko}}, \bibinfo {author} {\bibfnamefont {Y.}~\bibnamefont {Ishida}}, \bibinfo {author} {\bibfnamefont {X.}~\bibnamefont {Yang}}, \bibinfo {author} {\bibfnamefont {R.}~\bibnamefont {Kumai}}, \bibinfo {author} {\bibfnamefont {A.}~\bibnamefont {Toschi}}, \bibinfo {author} {\bibfnamefont {Y.}~\bibnamefont {Onose}}, \bibinfo {author} {\bibfnamefont {R.}~\bibnamefont {Arita}}, \bibinfo {author} {\bibfnamefont {K.}~\bibnamefont {Held}}, \bibinfo {author} {\bibfnamefont {O.~K.}\ \bibnamefont {Andersen}}, \bibinfo {author} {\bibfnamefont {S.}~\bibnamefont {Shin}},\ and\ \bibinfo {author} {\bibfnamefont {Y.}~\bibnamefont {Tokura}},\ }\bibfield  {title} {\bibinfo {title} {Pseudogap of metallic layered nickelate
  ${R}_{2\ensuremath{-}x}{\mathrm{sr}}_{x}{\mathrm{nio}}_{4}$ ($r=\mathrm{Nd},\mathrm{Eu}$) crystals measured using angle-resolved photoemission spectroscopy},\ }\href {https://doi.org/10.1103/PhysRevLett.106.027001} {\bibfield  {journal} {\bibinfo  {journal} {Phys. Rev. Lett.}\ }\textbf {\bibinfo {volume} {106}},\ \bibinfo {pages} {027001} (\bibinfo {year} {2011})}\BibitemShut {NoStop}%
\bibitem [{\citenamefont {Li}\ \emph {et~al.}(2019)\citenamefont {Li}, \citenamefont {Lee}, \citenamefont {Wang}, \citenamefont {Osada}, \citenamefont {Crossley}, \citenamefont {Lee}, \citenamefont {Cui}, \citenamefont {Hikita},\ and\ \citenamefont {Hwang}}]{li2019superconductivity}%
  \BibitemOpen
  \bibfield  {author} {\bibinfo {author} {\bibfnamefont {D.}~\bibnamefont {Li}}, \bibinfo {author} {\bibfnamefont {K.}~\bibnamefont {Lee}}, \bibinfo {author} {\bibfnamefont {B.~Y.}\ \bibnamefont {Wang}}, \bibinfo {author} {\bibfnamefont {M.}~\bibnamefont {Osada}}, \bibinfo {author} {\bibfnamefont {S.}~\bibnamefont {Crossley}}, \bibinfo {author} {\bibfnamefont {H.~R.}\ \bibnamefont {Lee}}, \bibinfo {author} {\bibfnamefont {Y.}~\bibnamefont {Cui}}, \bibinfo {author} {\bibfnamefont {Y.}~\bibnamefont {Hikita}},\ and\ \bibinfo {author} {\bibfnamefont {H.~Y.}\ \bibnamefont {Hwang}},\ }\bibfield  {title} {\bibinfo {title} {{Superconductivity in an infinite-layer nickelate}},\ }\href {https://doi.org/10.1038/s41586-019-1496-5} {\bibfield  {journal} {\bibinfo  {journal} {Nature}\ }\textbf {\bibinfo {volume} {572}},\ \bibinfo {pages} {624} (\bibinfo {year} {2019})}\BibitemShut {NoStop}%
\bibitem [{\citenamefont {Li}\ \emph {et~al.}(2020)\citenamefont {Li}, \citenamefont {Wang}, \citenamefont {Lee}, \citenamefont {Harvey}, \citenamefont {Osada}, \citenamefont {Goodge}, \citenamefont {Kourkoutis},\ and\ \citenamefont {Hwang}}]{Li2020}%
  \BibitemOpen
  \bibfield  {author} {\bibinfo {author} {\bibfnamefont {D.}~\bibnamefont {Li}}, \bibinfo {author} {\bibfnamefont {B.~Y.}\ \bibnamefont {Wang}}, \bibinfo {author} {\bibfnamefont {K.}~\bibnamefont {Lee}}, \bibinfo {author} {\bibfnamefont {S.~P.}\ \bibnamefont {Harvey}}, \bibinfo {author} {\bibfnamefont {M.}~\bibnamefont {Osada}}, \bibinfo {author} {\bibfnamefont {B.~H.}\ \bibnamefont {Goodge}}, \bibinfo {author} {\bibfnamefont {L.~F.}\ \bibnamefont {Kourkoutis}},\ and\ \bibinfo {author} {\bibfnamefont {H.~Y.}\ \bibnamefont {Hwang}},\ }\bibfield  {title} {\bibinfo {title} {{Superconducting Dome in {${\mathrm{Nd}}_{1\ensuremath{-}x}{\mathrm{Sr}}_{x}{\mathrm{NiO}}_{2}$} Infinite Layer Films}},\ }\href {https://doi.org/10.1103/PhysRevLett.125.027001} {\bibfield  {journal} {\bibinfo  {journal} {Phys. Rev. Lett.}\ }\textbf {\bibinfo {volume} {125}},\ \bibinfo {pages} {27001} (\bibinfo {year} {2020})}\BibitemShut {NoStop}%
\bibitem [{\citenamefont {Zeng}\ \emph {et~al.}(2020)\citenamefont {Zeng}, \citenamefont {Tang}, \citenamefont {Yin}, \citenamefont {Li}, \citenamefont {Li}, \citenamefont {Huang}, \citenamefont {Hu}, \citenamefont {Liu}, \citenamefont {Omar}, \citenamefont {Jani}, \citenamefont {Lim}, \citenamefont {Han}, \citenamefont {Wan}, \citenamefont {Yang}, \citenamefont {Pennycook}, \citenamefont {Wee},\ and\ \citenamefont {Ariando}}]{zeng2020}%
  \BibitemOpen
  \bibfield  {author} {\bibinfo {author} {\bibfnamefont {S.}~\bibnamefont {Zeng}}, \bibinfo {author} {\bibfnamefont {C.~S.}\ \bibnamefont {Tang}}, \bibinfo {author} {\bibfnamefont {X.}~\bibnamefont {Yin}}, \bibinfo {author} {\bibfnamefont {C.}~\bibnamefont {Li}}, \bibinfo {author} {\bibfnamefont {M.}~\bibnamefont {Li}}, \bibinfo {author} {\bibfnamefont {Z.}~\bibnamefont {Huang}}, \bibinfo {author} {\bibfnamefont {J.}~\bibnamefont {Hu}}, \bibinfo {author} {\bibfnamefont {W.}~\bibnamefont {Liu}}, \bibinfo {author} {\bibfnamefont {G.~J.}\ \bibnamefont {Omar}}, \bibinfo {author} {\bibfnamefont {H.}~\bibnamefont {Jani}}, \bibinfo {author} {\bibfnamefont {Z.~S.}\ \bibnamefont {Lim}}, \bibinfo {author} {\bibfnamefont {K.}~\bibnamefont {Han}}, \bibinfo {author} {\bibfnamefont {D.}~\bibnamefont {Wan}}, \bibinfo {author} {\bibfnamefont {P.}~\bibnamefont {Yang}}, \bibinfo {author} {\bibfnamefont {S.~J.}\ \bibnamefont {Pennycook}}, \bibinfo {author} {\bibfnamefont {A.~T.~S.}\ \bibnamefont {Wee}},\ and\ \bibinfo {author}
  {\bibfnamefont {A.}~\bibnamefont {Ariando}},\ }\bibfield  {title} {\bibinfo {title} {{Phase Diagram and Superconducting Dome of Infinite-Layer {${\mathrm{Nd}}_{1\ensuremath{-}x}{\mathrm{Sr}}_{x}{\mathrm{NiO}}_{2}$} Thin Films, arXiv:2004.11281}},\ }\href {https://doi.org/10.1103/PhysRevLett.125.147003} {\bibfield  {journal} {\bibinfo  {journal} {Phys. Rev. Lett.}\ }\textbf {\bibinfo {volume} {125}},\ \bibinfo {pages} {147003} (\bibinfo {year} {2020})}\BibitemShut {NoStop}%
\bibitem [{\citenamefont {Osada}\ \emph {et~al.}(2020)\citenamefont {Osada}, \citenamefont {Wang}, \citenamefont {Lee}, \citenamefont {Li},\ and\ \citenamefont {Hwang}}]{Osada2020}%
  \BibitemOpen
  \bibfield  {author} {\bibinfo {author} {\bibfnamefont {M.}~\bibnamefont {Osada}}, \bibinfo {author} {\bibfnamefont {B.~Y.}\ \bibnamefont {Wang}}, \bibinfo {author} {\bibfnamefont {K.}~\bibnamefont {Lee}}, \bibinfo {author} {\bibfnamefont {D.}~\bibnamefont {Li}},\ and\ \bibinfo {author} {\bibfnamefont {H.~Y.}\ \bibnamefont {Hwang}},\ }\bibfield  {title} {\bibinfo {title} {{Phase diagram of infinite layer praseodymium nickelate ${\mathrm{Pr}}_{1\ensuremath{-}x}{\mathrm{Sr}}_{x}{\mathrm{NiO}}_{2}$ thin films}},\ }\href {https://doi.org/10.1103/PhysRevMaterials.4.121801} {\bibfield  {journal} {\bibinfo  {journal} {Phys. Rev. Materials}\ }\textbf {\bibinfo {volume} {4}},\ \bibinfo {pages} {121801} (\bibinfo {year} {2020})}\BibitemShut {NoStop}%
\bibitem [{\citenamefont {Zeng}\ \emph {et~al.}(2022)\citenamefont {Zeng}, \citenamefont {Li}, \citenamefont {Chow}, \citenamefont {Cao}, \citenamefont {Zhang}, \citenamefont {Tang}, \citenamefont {Yin}, \citenamefont {Lim}, \citenamefont {Hu}, \citenamefont {Yang} \emph {et~al.}}]{zeng2022superconductivity}%
  \BibitemOpen
  \bibfield  {author} {\bibinfo {author} {\bibfnamefont {S.}~\bibnamefont {Zeng}}, \bibinfo {author} {\bibfnamefont {C.}~\bibnamefont {Li}}, \bibinfo {author} {\bibfnamefont {L.~E.}\ \bibnamefont {Chow}}, \bibinfo {author} {\bibfnamefont {Y.}~\bibnamefont {Cao}}, \bibinfo {author} {\bibfnamefont {Z.}~\bibnamefont {Zhang}}, \bibinfo {author} {\bibfnamefont {C.~S.}\ \bibnamefont {Tang}}, \bibinfo {author} {\bibfnamefont {X.}~\bibnamefont {Yin}}, \bibinfo {author} {\bibfnamefont {Z.~S.}\ \bibnamefont {Lim}}, \bibinfo {author} {\bibfnamefont {J.}~\bibnamefont {Hu}}, \bibinfo {author} {\bibfnamefont {P.}~\bibnamefont {Yang}}, \emph {et~al.},\ }\bibfield  {title} {\bibinfo {title} {Superconductivity in infinite-layer nickelate la1- xcaxnio2 thin films},\ }\href {https://doi.org/10.1126/sciadv.abl9927} {\bibfield  {journal} {\bibinfo  {journal} {Science advances}\ }\textbf {\bibinfo {volume} {8}},\ \bibinfo {pages} {eabl9927} (\bibinfo {year} {2022})}\BibitemShut {NoStop}%
\bibitem [{\citenamefont {Pan}\ \emph {et~al.}(2021)\citenamefont {Pan}, \citenamefont {Segedin}, \citenamefont {LaBollita}, \citenamefont {Song}, \citenamefont {Nica}, \citenamefont {Goodge}, \citenamefont {Pierce}, \citenamefont {Doyle}, \citenamefont {Novakov}, \citenamefont {Carrizales}, \citenamefont {N'Diaye}, \citenamefont {Shafer}, \citenamefont {Paik}, \citenamefont {Heron}, \citenamefont {Mason}, \citenamefont {Yacoby}, \citenamefont {Kourkoutis}, \citenamefont {Erten}, \citenamefont {Brooks}, \citenamefont {Botana},\ and\ \citenamefont {Mundy}}]{pan2021}%
  \BibitemOpen
  \bibfield  {author} {\bibinfo {author} {\bibfnamefont {G.~A.}\ \bibnamefont {Pan}}, \bibinfo {author} {\bibfnamefont {D.~F.}\ \bibnamefont {Segedin}}, \bibinfo {author} {\bibfnamefont {H.}~\bibnamefont {LaBollita}}, \bibinfo {author} {\bibfnamefont {Q.}~\bibnamefont {Song}}, \bibinfo {author} {\bibfnamefont {E.~M.}\ \bibnamefont {Nica}}, \bibinfo {author} {\bibfnamefont {B.~H.}\ \bibnamefont {Goodge}}, \bibinfo {author} {\bibfnamefont {A.~T.}\ \bibnamefont {Pierce}}, \bibinfo {author} {\bibfnamefont {S.}~\bibnamefont {Doyle}}, \bibinfo {author} {\bibfnamefont {S.}~\bibnamefont {Novakov}}, \bibinfo {author} {\bibfnamefont {D.~C.}\ \bibnamefont {Carrizales}}, \bibinfo {author} {\bibfnamefont {A.~T.}\ \bibnamefont {N'Diaye}}, \bibinfo {author} {\bibfnamefont {P.}~\bibnamefont {Shafer}}, \bibinfo {author} {\bibfnamefont {H.}~\bibnamefont {Paik}}, \bibinfo {author} {\bibfnamefont {J.~T.}\ \bibnamefont {Heron}}, \bibinfo {author} {\bibfnamefont {J.~A.}\ \bibnamefont {Mason}}, \bibinfo {author} {\bibfnamefont
  {A.}~\bibnamefont {Yacoby}}, \bibinfo {author} {\bibfnamefont {L.~F.}\ \bibnamefont {Kourkoutis}}, \bibinfo {author} {\bibfnamefont {O.}~\bibnamefont {Erten}}, \bibinfo {author} {\bibfnamefont {C.~M.}\ \bibnamefont {Brooks}}, \bibinfo {author} {\bibfnamefont {A.~S.}\ \bibnamefont {Botana}},\ and\ \bibinfo {author} {\bibfnamefont {J.~A.}\ \bibnamefont {Mundy}},\ }\bibfield  {title} {\bibinfo {title} {{Superconductivity in a quintuple-layer square-planar nickelate}},\ }\href {https://doi.org/10.1038/s41563-021-01142-9} {\bibfield  {journal} {\bibinfo  {journal} {Nature Materials}\ }\textbf {\bibinfo {volume} {21}},\ \bibinfo {pages} {160} (\bibinfo {year} {2021})}\BibitemShut {NoStop}%
\bibitem [{\citenamefont {Osada}\ \emph {et~al.}(2021)\citenamefont {Osada}, \citenamefont {Wang}, \citenamefont {Goodge}, \citenamefont {Harvey}, \citenamefont {Lee}, \citenamefont {Li}, \citenamefont {Kourkoutis},\ and\ \citenamefont {Hwang}}]{Osada2021}%
  \BibitemOpen
  \bibfield  {author} {\bibinfo {author} {\bibfnamefont {M.}~\bibnamefont {Osada}}, \bibinfo {author} {\bibfnamefont {B.~Y.}\ \bibnamefont {Wang}}, \bibinfo {author} {\bibfnamefont {B.~H.}\ \bibnamefont {Goodge}}, \bibinfo {author} {\bibfnamefont {S.~P.}\ \bibnamefont {Harvey}}, \bibinfo {author} {\bibfnamefont {K.}~\bibnamefont {Lee}}, \bibinfo {author} {\bibfnamefont {D.}~\bibnamefont {Li}}, \bibinfo {author} {\bibfnamefont {L.~F.}\ \bibnamefont {Kourkoutis}},\ and\ \bibinfo {author} {\bibfnamefont {H.~Y.}\ \bibnamefont {Hwang}},\ }\bibfield  {title} {\bibinfo {title} {{Nickelate Superconductivity without Rare-Earth Magnetism: {${\mathrm{(La,Sr)NiO}}_{2}$}}},\ }\href {https://doi.org/10.1002/adma.202104083} {\bibfield  {journal} {\bibinfo  {journal} {Advanced Materials}\ ,\ \bibinfo {pages} {2104083}} (\bibinfo {year} {2021})}\BibitemShut {NoStop}%
\bibitem [{\citenamefont {Wang}\ \emph {et~al.}(2022)\citenamefont {Wang}, \citenamefont {Yang}, \citenamefont {Yang}, \citenamefont {Chen}, \citenamefont {Zhang}, \citenamefont {Zhang}, \citenamefont {Zhu}, \citenamefont {Uwatoko}, \citenamefont {Gu}, \citenamefont {Dong}, \citenamefont {Sun}, \citenamefont {Jin},\ and\ \citenamefont {Cheng}}]{Wang2022}%
  \BibitemOpen
  \bibfield  {author} {\bibinfo {author} {\bibfnamefont {N.~N.}\ \bibnamefont {Wang}}, \bibinfo {author} {\bibfnamefont {M.~W.}\ \bibnamefont {Yang}}, \bibinfo {author} {\bibfnamefont {Z.}~\bibnamefont {Yang}}, \bibinfo {author} {\bibfnamefont {K.~Y.}\ \bibnamefont {Chen}}, \bibinfo {author} {\bibfnamefont {H.}~\bibnamefont {Zhang}}, \bibinfo {author} {\bibfnamefont {Q.~H.}\ \bibnamefont {Zhang}}, \bibinfo {author} {\bibfnamefont {Z.~H.}\ \bibnamefont {Zhu}}, \bibinfo {author} {\bibfnamefont {Y.}~\bibnamefont {Uwatoko}}, \bibinfo {author} {\bibfnamefont {L.}~\bibnamefont {Gu}}, \bibinfo {author} {\bibfnamefont {X.~L.}\ \bibnamefont {Dong}}, \bibinfo {author} {\bibfnamefont {J.~P.}\ \bibnamefont {Sun}}, \bibinfo {author} {\bibfnamefont {K.~J.}\ \bibnamefont {Jin}},\ and\ \bibinfo {author} {\bibfnamefont {J.-G.}\ \bibnamefont {Cheng}},\ }\bibfield  {title} {\bibinfo {title} {Pressure-induced monotonic enhancement of tc to over 30{\thinspace}{K} in superconducting {Pr$_{0.82}$Sr$_{0.18}$NiO$_2$} thin films},\
  }\href {https://doi.org/10.1038/s41467-022-32065-x} {\bibfield  {journal} {\bibinfo  {journal} {Nature Communications}\ }\textbf {\bibinfo {volume} {13}},\ \bibinfo {pages} {4367} (\bibinfo {year} {2022})}\BibitemShut {NoStop}%
\bibitem [{\citenamefont {Botana}\ and\ \citenamefont {Norman}(2020)}]{Botana2019}%
  \BibitemOpen
  \bibfield  {author} {\bibinfo {author} {\bibfnamefont {A.~S.}\ \bibnamefont {Botana}}\ and\ \bibinfo {author} {\bibfnamefont {M.~R.}\ \bibnamefont {Norman}},\ }\bibfield  {title} {\bibinfo {title} {{Similarities and Differences between {${\mathrm{LaNiO}}_{2}$} and {${\mathrm{CaCuO}}_{2}$} and Implications for Superconductivity}},\ }\href {https://doi.org/10.1103/PhysRevX.10.011024} {\bibfield  {journal} {\bibinfo  {journal} {Phys. Rev. X}\ }\textbf {\bibinfo {volume} {10}},\ \bibinfo {pages} {11024} (\bibinfo {year} {2020})}\BibitemShut {NoStop}%
\bibitem [{\citenamefont {Sakakibara}\ \emph {et~al.}(2020)\citenamefont {Sakakibara}, \citenamefont {Usui}, \citenamefont {Suzuki}, \citenamefont {Kotani}, \citenamefont {Aoki},\ and\ \citenamefont {Kuroki}}]{Hirofumi2019}%
  \BibitemOpen
  \bibfield  {author} {\bibinfo {author} {\bibfnamefont {H.}~\bibnamefont {Sakakibara}}, \bibinfo {author} {\bibfnamefont {H.}~\bibnamefont {Usui}}, \bibinfo {author} {\bibfnamefont {K.}~\bibnamefont {Suzuki}}, \bibinfo {author} {\bibfnamefont {T.}~\bibnamefont {Kotani}}, \bibinfo {author} {\bibfnamefont {H.}~\bibnamefont {Aoki}},\ and\ \bibinfo {author} {\bibfnamefont {K.}~\bibnamefont {Kuroki}},\ }\bibfield  {title} {\bibinfo {title} {{Model Construction and a Possibility of Cupratelike Pairing in a New ${d}^{9}$ Nickelate Superconductor {$(\mathrm{Nd},\mathrm{Sr}){\mathrm{NiO}}_{2}$}}},\ }\href {https://doi.org/10.1103/PhysRevLett.125.077003} {\bibfield  {journal} {\bibinfo  {journal} {Phys. Rev. Lett.}\ }\textbf {\bibinfo {volume} {125}},\ \bibinfo {pages} {77003} (\bibinfo {year} {2020})}\BibitemShut {NoStop}%
\bibitem [{\citenamefont {Jiang}\ \emph {et~al.}(2019)\citenamefont {Jiang}, \citenamefont {Si}, \citenamefont {Liao},\ and\ \citenamefont {Zhong}}]{jiang2019electronic}%
  \BibitemOpen
  \bibfield  {author} {\bibinfo {author} {\bibfnamefont {P.}~\bibnamefont {Jiang}}, \bibinfo {author} {\bibfnamefont {L.}~\bibnamefont {Si}}, \bibinfo {author} {\bibfnamefont {Z.}~\bibnamefont {Liao}},\ and\ \bibinfo {author} {\bibfnamefont {Z.}~\bibnamefont {Zhong}},\ }\bibfield  {title} {\bibinfo {title} {{Electronic structure of rare-earth infinite-layer ${R}\mathrm{Ni}\mathrm{O}_{2}$ $({R}=\mathrm{La},\mathrm{Nd})$}},\ }\href {https://doi.org/10.1103/PhysRevB.100.201106} {\bibfield  {journal} {\bibinfo  {journal} {Phys. Rev. B}\ }\textbf {\bibinfo {volume} {100}},\ \bibinfo {pages} {201106} (\bibinfo {year} {2019})}\BibitemShut {NoStop}%
\bibitem [{\citenamefont {Hirayama}\ \emph {et~al.}(2020)\citenamefont {Hirayama}, \citenamefont {Tadano}, \citenamefont {Nomura},\ and\ \citenamefont {Arita}}]{Motoaki2019}%
  \BibitemOpen
  \bibfield  {author} {\bibinfo {author} {\bibfnamefont {M.}~\bibnamefont {Hirayama}}, \bibinfo {author} {\bibfnamefont {T.}~\bibnamefont {Tadano}}, \bibinfo {author} {\bibfnamefont {Y.}~\bibnamefont {Nomura}},\ and\ \bibinfo {author} {\bibfnamefont {R.}~\bibnamefont {Arita}},\ }\bibfield  {title} {\bibinfo {title} {{Materials design of dynamically stable ${d}^{9}$ layered nickelates}},\ }\href {https://doi.org/10.1103/PhysRevB.101.075107} {\bibfield  {journal} {\bibinfo  {journal} {Phys. Rev. B}\ }\textbf {\bibinfo {volume} {101}},\ \bibinfo {pages} {75107} (\bibinfo {year} {2020})}\BibitemShut {NoStop}%
\bibitem [{\citenamefont {Hu}\ and\ \citenamefont {Wu}(2019)}]{hu2019twoband}%
  \BibitemOpen
  \bibfield  {author} {\bibinfo {author} {\bibfnamefont {L.-H.}\ \bibnamefont {Hu}}\ and\ \bibinfo {author} {\bibfnamefont {C.}~\bibnamefont {Wu}},\ }\bibfield  {title} {\bibinfo {title} {{Two-band model for magnetism and superconductivity in nickelates}},\ }\href {https://doi.org/10.1103/PhysRevResearch.1.032046} {\bibfield  {journal} {\bibinfo  {journal} {Phys. Rev. Research}\ }\textbf {\bibinfo {volume} {1}},\ \bibinfo {pages} {32046} (\bibinfo {year} {2019})}\BibitemShut {NoStop}%
\bibitem [{\citenamefont {Wu}\ \emph {et~al.}(2020)\citenamefont {Wu}, \citenamefont {{Di Sante}}, \citenamefont {Schwemmer}, \citenamefont {Hanke}, \citenamefont {Hwang}, \citenamefont {Raghu},\ and\ \citenamefont {Thomale}}]{Wu2019}%
  \BibitemOpen
  \bibfield  {author} {\bibinfo {author} {\bibfnamefont {X.}~\bibnamefont {Wu}}, \bibinfo {author} {\bibfnamefont {D.}~\bibnamefont {{Di Sante}}}, \bibinfo {author} {\bibfnamefont {T.}~\bibnamefont {Schwemmer}}, \bibinfo {author} {\bibfnamefont {W.}~\bibnamefont {Hanke}}, \bibinfo {author} {\bibfnamefont {H.~Y.}\ \bibnamefont {Hwang}}, \bibinfo {author} {\bibfnamefont {S.}~\bibnamefont {Raghu}},\ and\ \bibinfo {author} {\bibfnamefont {R.}~\bibnamefont {Thomale}},\ }\bibfield  {title} {\bibinfo {title} {{Robust ${d}_{{x}^{2}\ensuremath{-}{y}^{2}}$-wave superconductivity of infinite-layer nickelates}},\ }\href {https://doi.org/10.1103/PhysRevB.101.060504} {\bibfield  {journal} {\bibinfo  {journal} {Phys. Rev. B}\ }\textbf {\bibinfo {volume} {101}},\ \bibinfo {pages} {60504} (\bibinfo {year} {2020})}\BibitemShut {NoStop}%
\bibitem [{\citenamefont {Nomura}\ \emph {et~al.}(2019)\citenamefont {Nomura}, \citenamefont {Hirayama}, \citenamefont {Tadano}, \citenamefont {Yoshimoto}, \citenamefont {Nakamura},\ and\ \citenamefont {Arita}}]{Nomura2019}%
  \BibitemOpen
  \bibfield  {author} {\bibinfo {author} {\bibfnamefont {Y.}~\bibnamefont {Nomura}}, \bibinfo {author} {\bibfnamefont {M.}~\bibnamefont {Hirayama}}, \bibinfo {author} {\bibfnamefont {T.}~\bibnamefont {Tadano}}, \bibinfo {author} {\bibfnamefont {Y.}~\bibnamefont {Yoshimoto}}, \bibinfo {author} {\bibfnamefont {K.}~\bibnamefont {Nakamura}},\ and\ \bibinfo {author} {\bibfnamefont {R.}~\bibnamefont {Arita}},\ }\bibfield  {title} {\bibinfo {title} {{Formation of a two-dimensional single-component correlated electron system and band engineering in the nickelate superconductor {${\mathrm{NdNiO}}_{2}$}}},\ }\href {https://doi.org/10.1103/PhysRevB.100.205138} {\bibfield  {journal} {\bibinfo  {journal} {Phys. Rev. B}\ }\textbf {\bibinfo {volume} {100}},\ \bibinfo {pages} {205138} (\bibinfo {year} {2019})}\BibitemShut {NoStop}%
\bibitem [{\citenamefont {Zhang}\ \emph {et~al.}(2020)\citenamefont {Zhang}, \citenamefont {Yang},\ and\ \citenamefont {Zhang}}]{Zhang2019}%
  \BibitemOpen
  \bibfield  {author} {\bibinfo {author} {\bibfnamefont {G.-M.}\ \bibnamefont {Zhang}}, \bibinfo {author} {\bibfnamefont {Y.-F.}\ \bibnamefont {Yang}},\ and\ \bibinfo {author} {\bibfnamefont {F.-C.}\ \bibnamefont {Zhang}},\ }\bibfield  {title} {\bibinfo {title} {{Self-doped {Mott} insulator for parent compounds of nickelate superconductors}},\ }\href {https://doi.org/10.1103/PhysRevB.101.020501} {\bibfield  {journal} {\bibinfo  {journal} {Phys. Rev. B}\ }\textbf {\bibinfo {volume} {101}},\ \bibinfo {pages} {20501} (\bibinfo {year} {2020})}\BibitemShut {NoStop}%
\bibitem [{\citenamefont {Jiang}\ \emph {et~al.}(2020)\citenamefont {Jiang}, \citenamefont {Berciu},\ and\ \citenamefont {Sawatzky}}]{Jiang2019}%
  \BibitemOpen
  \bibfield  {author} {\bibinfo {author} {\bibfnamefont {M.}~\bibnamefont {Jiang}}, \bibinfo {author} {\bibfnamefont {M.}~\bibnamefont {Berciu}},\ and\ \bibinfo {author} {\bibfnamefont {G.~A.}\ \bibnamefont {Sawatzky}},\ }\bibfield  {title} {\bibinfo {title} {{Critical Nature of the {Ni} Spin State in Doped {${\mathrm{NdNiO}}_{2}$}}},\ }\href {https://doi.org/10.1103/PhysRevLett.124.207004} {\bibfield  {journal} {\bibinfo  {journal} {Phys. Rev. Lett.}\ }\textbf {\bibinfo {volume} {124}},\ \bibinfo {pages} {207004} (\bibinfo {year} {2020})}\BibitemShut {NoStop}%
\bibitem [{\citenamefont {Werner}\ and\ \citenamefont {Hoshino}(2020)}]{Werner2019}%
  \BibitemOpen
  \bibfield  {author} {\bibinfo {author} {\bibfnamefont {P.}~\bibnamefont {Werner}}\ and\ \bibinfo {author} {\bibfnamefont {S.}~\bibnamefont {Hoshino}},\ }\bibfield  {title} {\bibinfo {title} {{Nickelate superconductors: Multiorbital nature and spin freezing}},\ }\href {https://doi.org/10.1103/PhysRevB.101.041104} {\bibfield  {journal} {\bibinfo  {journal} {Phys. Rev. B}\ }\textbf {\bibinfo {volume} {101}},\ \bibinfo {pages} {41104} (\bibinfo {year} {2020})}\BibitemShut {NoStop}%
\bibitem [{\citenamefont {Si}\ \emph {et~al.}(2020)\citenamefont {Si}, \citenamefont {Xiao}, \citenamefont {Kaufmann}, \citenamefont {Tomczak}, \citenamefont {Lu}, \citenamefont {Zhong},\ and\ \citenamefont {Held}}]{Si2020}%
  \BibitemOpen
  \bibfield  {author} {\bibinfo {author} {\bibfnamefont {L.}~\bibnamefont {Si}}, \bibinfo {author} {\bibfnamefont {W.}~\bibnamefont {Xiao}}, \bibinfo {author} {\bibfnamefont {J.}~\bibnamefont {Kaufmann}}, \bibinfo {author} {\bibfnamefont {J.~M.}\ \bibnamefont {Tomczak}}, \bibinfo {author} {\bibfnamefont {Y.}~\bibnamefont {Lu}}, \bibinfo {author} {\bibfnamefont {Z.}~\bibnamefont {Zhong}},\ and\ \bibinfo {author} {\bibfnamefont {K.}~\bibnamefont {Held}},\ }\bibfield  {title} {\bibinfo {title} {{Topotactic Hydrogen in Nickelate Superconductors and Akin Infinite-Layer Oxides {$AB{\mathrm{O}}_{2}$}}},\ }\href {https://doi.org/10.1103/PhysRevLett.124.166402} {\bibfield  {journal} {\bibinfo  {journal} {Phys. Rev. Lett.}\ }\textbf {\bibinfo {volume} {124}},\ \bibinfo {pages} {166402} (\bibinfo {year} {2020})}\BibitemShut {NoStop}%
\bibitem [{\citenamefont {Nomura}\ and\ \citenamefont {Arita}(2022)}]{Nomura2022}%
  \BibitemOpen
  \bibfield  {author} {\bibinfo {author} {\bibfnamefont {Y.}~\bibnamefont {Nomura}}\ and\ \bibinfo {author} {\bibfnamefont {R.}~\bibnamefont {Arita}},\ }\bibfield  {title} {\bibinfo {title} {Superconductivity in infinite-layer nickelates},\ }\href {https://doi.org/10.1088/1361-6633/ac5a60} {\bibfield  {journal} {\bibinfo  {journal} {Reports on Progress in Physics}\ }\textbf {\bibinfo {volume} {85}},\ \bibinfo {pages} {052501} (\bibinfo {year} {2022})}\BibitemShut {NoStop}%
\bibitem [{\citenamefont {Kitatani}\ \emph {et~al.}(2023)\citenamefont {Kitatani}, \citenamefont {Si}, \citenamefont {Worm}, \citenamefont {Tomczak}, \citenamefont {Arita},\ and\ \citenamefont {Held}}]{Kitatani2023b}%
  \BibitemOpen
  \bibfield  {author} {\bibinfo {author} {\bibfnamefont {M.}~\bibnamefont {Kitatani}}, \bibinfo {author} {\bibfnamefont {L.}~\bibnamefont {Si}}, \bibinfo {author} {\bibfnamefont {P.}~\bibnamefont {Worm}}, \bibinfo {author} {\bibfnamefont {J.~M.}\ \bibnamefont {Tomczak}}, \bibinfo {author} {\bibfnamefont {R.}~\bibnamefont {Arita}},\ and\ \bibinfo {author} {\bibfnamefont {K.}~\bibnamefont {Held}},\ }\bibfield  {title} {\bibinfo {title} {Optimizing superconductivity: From cuprates via nickelates to palladates},\ }\href {https://doi.org/10.1103/PhysRevLett.130.166002} {\bibfield  {journal} {\bibinfo  {journal} {Phys. Rev. Lett.}\ }\textbf {\bibinfo {volume} {130}},\ \bibinfo {pages} {166002} (\bibinfo {year} {2023})}\BibitemShut {NoStop}%
\bibitem [{\citenamefont {Lee}\ \emph {et~al.}(2023{\natexlab{a}})\citenamefont {Lee}, \citenamefont {Wang}, \citenamefont {Osada}, \citenamefont {Goodge}, \citenamefont {Wang}, \citenamefont {Lee}, \citenamefont {Harvey}, \citenamefont {Kim}, \citenamefont {Yu}, \citenamefont {Murthy} \emph {et~al.}}]{lee2023linear}%
  \BibitemOpen
  \bibfield  {author} {\bibinfo {author} {\bibfnamefont {K.}~\bibnamefont {Lee}}, \bibinfo {author} {\bibfnamefont {B.~Y.}\ \bibnamefont {Wang}}, \bibinfo {author} {\bibfnamefont {M.}~\bibnamefont {Osada}}, \bibinfo {author} {\bibfnamefont {B.~H.}\ \bibnamefont {Goodge}}, \bibinfo {author} {\bibfnamefont {T.~C.}\ \bibnamefont {Wang}}, \bibinfo {author} {\bibfnamefont {Y.}~\bibnamefont {Lee}}, \bibinfo {author} {\bibfnamefont {S.}~\bibnamefont {Harvey}}, \bibinfo {author} {\bibfnamefont {W.~J.}\ \bibnamefont {Kim}}, \bibinfo {author} {\bibfnamefont {Y.}~\bibnamefont {Yu}}, \bibinfo {author} {\bibfnamefont {C.}~\bibnamefont {Murthy}}, \emph {et~al.},\ }\bibfield  {title} {\bibinfo {title} {Linear-in-temperature resistivity for optimally superconducting (nd, sr) nio2},\ }\href {https://doi.org/doi.org/10.1038/s41586-023-06129-x} {\bibfield  {journal} {\bibinfo  {journal} {Nature}\ }\textbf {\bibinfo {volume} {619}},\ \bibinfo {pages} {288} (\bibinfo {year} {2023}{\natexlab{a}})}\BibitemShut {NoStop}%
\bibitem [{\citenamefont {Ando}\ \emph {et~al.}(2004)\citenamefont {Ando}, \citenamefont {Kurita}, \citenamefont {Komiya}, \citenamefont {Ono},\ and\ \citenamefont {Segawa}}]{PhysRevLett.92.197001}%
  \BibitemOpen
  \bibfield  {author} {\bibinfo {author} {\bibfnamefont {Y.}~\bibnamefont {Ando}}, \bibinfo {author} {\bibfnamefont {Y.}~\bibnamefont {Kurita}}, \bibinfo {author} {\bibfnamefont {S.}~\bibnamefont {Komiya}}, \bibinfo {author} {\bibfnamefont {S.}~\bibnamefont {Ono}},\ and\ \bibinfo {author} {\bibfnamefont {K.}~\bibnamefont {Segawa}},\ }\bibfield  {title} {\bibinfo {title} {Evolution of the hall coefficient and the peculiar electronic structure of the cuprate superconductors},\ }\href {https://doi.org/10.1103/PhysRevLett.92.197001} {\bibfield  {journal} {\bibinfo  {journal} {Phys. Rev. Lett.}\ }\textbf {\bibinfo {volume} {92}},\ \bibinfo {pages} {197001} (\bibinfo {year} {2004})}\BibitemShut {NoStop}%
\bibitem [{\citenamefont {Balakirev}\ \emph {et~al.}(2003)\citenamefont {Balakirev}, \citenamefont {Betts}, \citenamefont {Migliori}, \citenamefont {Ono}, \citenamefont {Ando},\ and\ \citenamefont {Boebinger}}]{balakirev2003signature}%
  \BibitemOpen
  \bibfield  {author} {\bibinfo {author} {\bibfnamefont {F.~F.}\ \bibnamefont {Balakirev}}, \bibinfo {author} {\bibfnamefont {J.~B.}\ \bibnamefont {Betts}}, \bibinfo {author} {\bibfnamefont {A.}~\bibnamefont {Migliori}}, \bibinfo {author} {\bibfnamefont {S.}~\bibnamefont {Ono}}, \bibinfo {author} {\bibfnamefont {Y.}~\bibnamefont {Ando}},\ and\ \bibinfo {author} {\bibfnamefont {G.~S.}\ \bibnamefont {Boebinger}},\ }\bibfield  {title} {\bibinfo {title} {Signature of optimal doping in hall-effect measurements on a high-temperature superconductor},\ }\href {https://doi.org/10.1038/nature01890} {\bibfield  {journal} {\bibinfo  {journal} {Nature}\ }\textbf {\bibinfo {volume} {424}},\ \bibinfo {pages} {912} (\bibinfo {year} {2003})}\BibitemShut {NoStop}%
\bibitem [{\citenamefont {Badoux}\ \emph {et~al.}(2016)\citenamefont {Badoux}, \citenamefont {Tabis}, \citenamefont {Lalibert{\'e}}, \citenamefont {Grissonnanche}, \citenamefont {Vignolle}, \citenamefont {Vignolles}, \citenamefont {B{\'e}ard}, \citenamefont {Bonn}, \citenamefont {Hardy}, \citenamefont {Liang} \emph {et~al.}}]{badoux2016change}%
  \BibitemOpen
  \bibfield  {author} {\bibinfo {author} {\bibfnamefont {S.}~\bibnamefont {Badoux}}, \bibinfo {author} {\bibfnamefont {W.}~\bibnamefont {Tabis}}, \bibinfo {author} {\bibfnamefont {F.}~\bibnamefont {Lalibert{\'e}}}, \bibinfo {author} {\bibfnamefont {G.}~\bibnamefont {Grissonnanche}}, \bibinfo {author} {\bibfnamefont {B.}~\bibnamefont {Vignolle}}, \bibinfo {author} {\bibfnamefont {D.}~\bibnamefont {Vignolles}}, \bibinfo {author} {\bibfnamefont {J.}~\bibnamefont {B{\'e}ard}}, \bibinfo {author} {\bibfnamefont {D.}~\bibnamefont {Bonn}}, \bibinfo {author} {\bibfnamefont {W.}~\bibnamefont {Hardy}}, \bibinfo {author} {\bibfnamefont {R.}~\bibnamefont {Liang}}, \emph {et~al.},\ }\bibfield  {title} {\bibinfo {title} {Change of carrier density at the pseudogap critical point of a cuprate superconductor},\ }\href {https://doi.org/10.1038/nature16983} {\bibfield  {journal} {\bibinfo  {journal} {Nature}\ }\textbf {\bibinfo {volume} {531}},\ \bibinfo {pages} {210} (\bibinfo {year} {2016})}\BibitemShut {NoStop}%
\bibitem [{\citenamefont {Metzner}\ and\ \citenamefont {Vollhardt}(1989)}]{Metzner1989}%
  \BibitemOpen
  \bibfield  {author} {\bibinfo {author} {\bibfnamefont {W.}~\bibnamefont {Metzner}}\ and\ \bibinfo {author} {\bibfnamefont {D.}~\bibnamefont {Vollhardt}},\ }\bibfield  {title} {\bibinfo {title} {{Correlated Lattice Fermions in $d=\infty$ Dimensions}},\ }\href {https://doi.org/10.1103/PhysRevLett.62.324} {\bibfield  {journal} {\bibinfo  {journal} {Phys. Rev. Lett.}\ }\textbf {\bibinfo {volume} {62}},\ \bibinfo {pages} {324} (\bibinfo {year} {1989})}\BibitemShut {NoStop}%
\bibitem [{\citenamefont {Georges}\ and\ \citenamefont {Kotliar}(1992)}]{Georges1992a}%
  \BibitemOpen
  \bibfield  {author} {\bibinfo {author} {\bibfnamefont {A.}~\bibnamefont {Georges}}\ and\ \bibinfo {author} {\bibfnamefont {G.}~\bibnamefont {Kotliar}},\ }\bibfield  {title} {\bibinfo {title} {{Hubbard model in infinite dimensions}},\ }\href {https://doi.org/10.1103/PhysRevB.45.6479} {\bibfield  {journal} {\bibinfo  {journal} {Phys. Rev. B}\ }\textbf {\bibinfo {volume} {45}},\ \bibinfo {pages} {6479} (\bibinfo {year} {1992})}\BibitemShut {NoStop}%
\bibitem [{\citenamefont {Georges}\ \emph {et~al.}(1996)\citenamefont {Georges}, \citenamefont {Kotliar}, \citenamefont {Krauth},\ and\ \citenamefont {Rozenberg}}]{Georges1996}%
  \BibitemOpen
  \bibfield  {author} {\bibinfo {author} {\bibfnamefont {A.}~\bibnamefont {Georges}}, \bibinfo {author} {\bibfnamefont {G.}~\bibnamefont {Kotliar}}, \bibinfo {author} {\bibfnamefont {W.}~\bibnamefont {Krauth}},\ and\ \bibinfo {author} {\bibfnamefont {M.~J.}\ \bibnamefont {Rozenberg}},\ }\bibfield  {title} {\bibinfo {title} {{Dynamical mean-field theory of strongly correlated fermion systems and the limit of infinite dimensions}},\ }\href {https://doi.org/10.1103/RevModPhys.68.13} {\bibfield  {journal} {\bibinfo  {journal} {Rev. Mod. Phys.}\ }\textbf {\bibinfo {volume} {68}},\ \bibinfo {pages} {13} (\bibinfo {year} {1996})}\BibitemShut {NoStop}%
\bibitem [{\citenamefont {Held}(2007)}]{held2007electronic}%
  \BibitemOpen
  \bibfield  {author} {\bibinfo {author} {\bibfnamefont {K.}~\bibnamefont {Held}},\ }\bibfield  {title} {\bibinfo {title} {{Electronic structure calculations using dynamical mean field theory}},\ }\href {https://doi.org/10.1080/00018730701619647} {\bibfield  {journal} {\bibinfo  {journal} {Advances in physics}\ }\textbf {\bibinfo {volume} {56}},\ \bibinfo {pages} {829} (\bibinfo {year} {2007})}\BibitemShut {NoStop}%
\bibitem [{\citenamefont {Kitatani}\ \emph {et~al.}(2020)\citenamefont {Kitatani}, \citenamefont {Si}, \citenamefont {Janson}, \citenamefont {Arita}, \citenamefont {Zhong},\ and\ \citenamefont {Held}}]{Kitatani2020}%
  \BibitemOpen
  \bibfield  {author} {\bibinfo {author} {\bibfnamefont {M.}~\bibnamefont {Kitatani}}, \bibinfo {author} {\bibfnamefont {L.}~\bibnamefont {Si}}, \bibinfo {author} {\bibfnamefont {O.}~\bibnamefont {Janson}}, \bibinfo {author} {\bibfnamefont {R.}~\bibnamefont {Arita}}, \bibinfo {author} {\bibfnamefont {Z.}~\bibnamefont {Zhong}},\ and\ \bibinfo {author} {\bibfnamefont {K.}~\bibnamefont {Held}},\ }\bibfield  {title} {\bibinfo {title} {{Nickelate superconductors -- a renaissance of the one-band Hubbard model}},\ }\href {https://doi.org/10.1038/s41535-020-00260-y} {\bibfield  {journal} {\bibinfo  {journal} {npj Quantum Materials}\ }\textbf {\bibinfo {volume} {5}},\ \bibinfo {pages} {59} (\bibinfo {year} {2020})}\BibitemShut {NoStop}%
\bibitem [{\citenamefont {Karp}\ \emph {et~al.}(2020)\citenamefont {Karp}, \citenamefont {Botana}, \citenamefont {Norman}, \citenamefont {Park}, \citenamefont {Zingl},\ and\ \citenamefont {Millis}}]{Karp2020}%
  \BibitemOpen
  \bibfield  {author} {\bibinfo {author} {\bibfnamefont {J.}~\bibnamefont {Karp}}, \bibinfo {author} {\bibfnamefont {A.~S.}\ \bibnamefont {Botana}}, \bibinfo {author} {\bibfnamefont {M.~R.}\ \bibnamefont {Norman}}, \bibinfo {author} {\bibfnamefont {H.}~\bibnamefont {Park}}, \bibinfo {author} {\bibfnamefont {M.}~\bibnamefont {Zingl}},\ and\ \bibinfo {author} {\bibfnamefont {A.}~\bibnamefont {Millis}},\ }\bibfield  {title} {\bibinfo {title} {{Many-Body Electronic Structure of {${\mathrm{NdNiO}}_{2}$} and {${\mathrm{CaCuO}}_{2}$}}},\ }\href {https://doi.org/10.1103/PhysRevX.10.021061} {\bibfield  {journal} {\bibinfo  {journal} {Phys. Rev. X}\ }\textbf {\bibinfo {volume} {10}},\ \bibinfo {pages} {21061} (\bibinfo {year} {2020})}\BibitemShut {NoStop}%
\bibitem [{\citenamefont {LaBollita}\ \emph {et~al.}(2022)\citenamefont {LaBollita}, \citenamefont {Jung},\ and\ \citenamefont {Botana}}]{LaBollita2022b}%
  \BibitemOpen
  \bibfield  {author} {\bibinfo {author} {\bibfnamefont {H.}~\bibnamefont {LaBollita}}, \bibinfo {author} {\bibfnamefont {M.-C.}\ \bibnamefont {Jung}},\ and\ \bibinfo {author} {\bibfnamefont {A.~S.}\ \bibnamefont {Botana}},\ }\bibfield  {title} {\bibinfo {title} {Many-body electronic structure of ${d}^{9\ensuremath{-}\ensuremath{\delta}}$ layered nickelates},\ }\href {https://doi.org/10.1103/PhysRevB.106.115132} {\bibfield  {journal} {\bibinfo  {journal} {Phys. Rev. B}\ }\textbf {\bibinfo {volume} {106}},\ \bibinfo {pages} {115132} (\bibinfo {year} {2022})}\BibitemShut {NoStop}%
\bibitem [{\citenamefont {Pascut}\ \emph {et~al.}(2023)\citenamefont {Pascut}, \citenamefont {Cosovanu}, \citenamefont {Haule},\ and\ \citenamefont {Quader}}]{Pascut2023}%
  \BibitemOpen
  \bibfield  {author} {\bibinfo {author} {\bibfnamefont {G.}~\bibnamefont {Pascut}}, \bibinfo {author} {\bibfnamefont {L.}~\bibnamefont {Cosovanu}}, \bibinfo {author} {\bibfnamefont {K.}~\bibnamefont {Haule}},\ and\ \bibinfo {author} {\bibfnamefont {K.~F.}\ \bibnamefont {Quader}},\ }\bibfield  {title} {\bibinfo {title} {Correlation-temperature phase diagram of prototypical infinite layer rare earth nickelates},\ }\href {https://doi.org/10.1038/s42005-023-01163-7} {\bibfield  {journal} {\bibinfo  {journal} {Commun. Phys.}\ }\textbf {\bibinfo {volume} {6}},\ \bibinfo {pages} {45} (\bibinfo {year} {2023})}\BibitemShut {NoStop}%
\bibitem [{\citenamefont {Held}\ \emph {et~al.}(2022)\citenamefont {Held}, \citenamefont {Si}, \citenamefont {Worm}, \citenamefont {Janson}, \citenamefont {Arita}, \citenamefont {Zhong}, \citenamefont {Tomczak},\ and\ \citenamefont {Kitatani}}]{Held2022}%
  \BibitemOpen
  \bibfield  {author} {\bibinfo {author} {\bibfnamefont {K.}~\bibnamefont {Held}}, \bibinfo {author} {\bibfnamefont {L.}~\bibnamefont {Si}}, \bibinfo {author} {\bibfnamefont {P.}~\bibnamefont {Worm}}, \bibinfo {author} {\bibfnamefont {O.}~\bibnamefont {Janson}}, \bibinfo {author} {\bibfnamefont {R.}~\bibnamefont {Arita}}, \bibinfo {author} {\bibfnamefont {Z.}~\bibnamefont {Zhong}}, \bibinfo {author} {\bibfnamefont {J.~M.}\ \bibnamefont {Tomczak}},\ and\ \bibinfo {author} {\bibfnamefont {M.}~\bibnamefont {Kitatani}},\ }\bibfield  {title} {\bibinfo {title} {{Phase Diagram of Nickelate Superconductors Calculated by Dynamical Vertex Approximation}},\ }\href {https://doi.org/10.3389/fphy.2021.810394} {\bibfield  {journal} {\bibinfo  {journal} {Frontiers in Physics}\ }\textbf {\bibinfo {volume} {9}},\ \bibinfo {pages} {810394} (\bibinfo {year} {2022})}\BibitemShut {NoStop}%
\bibitem [{\citenamefont {Toschi}\ \emph {et~al.}(2007)\citenamefont {Toschi}, \citenamefont {Katanin},\ and\ \citenamefont {Held}}]{Toschi2007}%
  \BibitemOpen
  \bibfield  {author} {\bibinfo {author} {\bibfnamefont {A.}~\bibnamefont {Toschi}}, \bibinfo {author} {\bibfnamefont {A.~A.}\ \bibnamefont {Katanin}},\ and\ \bibinfo {author} {\bibfnamefont {K.}~\bibnamefont {Held}},\ }\bibfield  {title} {\bibinfo {title} {{Dynamical vertex approximation; A step beyond dynamical mean-field theory}},\ }\href {https://doi.org/10.1103/PhysRevB.75.045118} {\bibfield  {journal} {\bibinfo  {journal} {Phys Rev. B}\ }\textbf {\bibinfo {volume} {75}},\ \bibinfo {pages} {45118} (\bibinfo {year} {2007})}\BibitemShut {NoStop}%
\bibitem [{\citenamefont {Rohringer}\ \emph {et~al.}(2018)\citenamefont {Rohringer}, \citenamefont {Hafermann}, \citenamefont {Toschi}, \citenamefont {Katanin}, \citenamefont {Antipov}, \citenamefont {Katsnelson}, \citenamefont {Lichtenstein}, \citenamefont {Rubtsov},\ and\ \citenamefont {Held}}]{RMPVertex}%
  \BibitemOpen
  \bibfield  {author} {\bibinfo {author} {\bibfnamefont {G.}~\bibnamefont {Rohringer}}, \bibinfo {author} {\bibfnamefont {H.}~\bibnamefont {Hafermann}}, \bibinfo {author} {\bibfnamefont {A.}~\bibnamefont {Toschi}}, \bibinfo {author} {\bibfnamefont {A.~A.}\ \bibnamefont {Katanin}}, \bibinfo {author} {\bibfnamefont {A.~E.}\ \bibnamefont {Antipov}}, \bibinfo {author} {\bibfnamefont {M.~I.}\ \bibnamefont {Katsnelson}}, \bibinfo {author} {\bibfnamefont {A.~I.}\ \bibnamefont {Lichtenstein}}, \bibinfo {author} {\bibfnamefont {A.~N.}\ \bibnamefont {Rubtsov}},\ and\ \bibinfo {author} {\bibfnamefont {K.}~\bibnamefont {Held}},\ }\bibfield  {title} {\bibinfo {title} {{Diagrammatic routes to nonlocal correlations beyond dynamical mean field theory}},\ }\href {https://doi.org/10.1103/RevModPhys.90.025003} {\bibfield  {journal} {\bibinfo  {journal} {Rev. Mod. Phys.}\ }\textbf {\bibinfo {volume} {90}},\ \bibinfo {pages} {25003} (\bibinfo {year} {2018})}\BibitemShut {NoStop}%
\bibitem [{\citenamefont {Lee}\ \emph {et~al.}(2023{\natexlab{b}})\citenamefont {Lee}, \citenamefont {Wang}, \citenamefont {Osada}, \citenamefont {Goodge}, \citenamefont {Wang}, \citenamefont {Lee}, \citenamefont {Harvey}, \citenamefont {Kim}, \citenamefont {Yu}, \citenamefont {Murthy}, \citenamefont {Raghu}, \citenamefont {Kourkoutis},\ and\ \citenamefont {Hwang}}]{Lee2023}%
  \BibitemOpen
  \bibfield  {author} {\bibinfo {author} {\bibfnamefont {K.}~\bibnamefont {Lee}}, \bibinfo {author} {\bibfnamefont {B.~Y.}\ \bibnamefont {Wang}}, \bibinfo {author} {\bibfnamefont {M.}~\bibnamefont {Osada}}, \bibinfo {author} {\bibfnamefont {B.~H.}\ \bibnamefont {Goodge}}, \bibinfo {author} {\bibfnamefont {T.~C.}\ \bibnamefont {Wang}}, \bibinfo {author} {\bibfnamefont {Y.}~\bibnamefont {Lee}}, \bibinfo {author} {\bibfnamefont {S.}~\bibnamefont {Harvey}}, \bibinfo {author} {\bibfnamefont {W.~J.}\ \bibnamefont {Kim}}, \bibinfo {author} {\bibfnamefont {Y.}~\bibnamefont {Yu}}, \bibinfo {author} {\bibfnamefont {C.}~\bibnamefont {Murthy}}, \bibinfo {author} {\bibfnamefont {S.}~\bibnamefont {Raghu}}, \bibinfo {author} {\bibfnamefont {L.~F.}\ \bibnamefont {Kourkoutis}},\ and\ \bibinfo {author} {\bibfnamefont {H.~Y.}\ \bibnamefont {Hwang}},\ }\bibfield  {title} {\bibinfo {title} {Linear-in-temperature resistivity for optimally superconducting ({Nd}, {Sr}){NiO$_2$}},\ }\href
  {https://www.nature.com/articles/s41586-023-06129-x} {\bibfield  {journal} {\bibinfo  {journal} {Nature}\ }\textbf {\bibinfo {volume} {619}},\ \bibinfo {pages} {288} (\bibinfo {year} {2023}{\natexlab{b}})}\BibitemShut {NoStop}%
\bibitem [{Note1()}]{Note1}%
  \BibitemOpen
  \bibinfo {note} {For a one-to-one comparison, see \cite {Worm2024}}\BibitemShut {NoStop}%
\bibitem [{\citenamefont {Lu}\ \emph {et~al.}(2021)\citenamefont {Lu}, \citenamefont {Rossi}, \citenamefont {Nag}, \citenamefont {Osada}, \citenamefont {Li}, \citenamefont {Lee}, \citenamefont {Wang}, \citenamefont {Garcia-Fernandez}, \citenamefont {Agrestini}, \citenamefont {Shen}, \citenamefont {Been}, \citenamefont {Moritz}, \citenamefont {Devereaux}, \citenamefont {Zaanen}, \citenamefont {Hwang}, \citenamefont {Zhou},\ and\ \citenamefont {Lee}}]{Lu2021}%
  \BibitemOpen
  \bibfield  {author} {\bibinfo {author} {\bibfnamefont {H.}~\bibnamefont {Lu}}, \bibinfo {author} {\bibfnamefont {M.}~\bibnamefont {Rossi}}, \bibinfo {author} {\bibfnamefont {A.}~\bibnamefont {Nag}}, \bibinfo {author} {\bibfnamefont {M.}~\bibnamefont {Osada}}, \bibinfo {author} {\bibfnamefont {D.~F.}\ \bibnamefont {Li}}, \bibinfo {author} {\bibfnamefont {K.}~\bibnamefont {Lee}}, \bibinfo {author} {\bibfnamefont {B.~Y.}\ \bibnamefont {Wang}}, \bibinfo {author} {\bibfnamefont {M.}~\bibnamefont {Garcia-Fernandez}}, \bibinfo {author} {\bibfnamefont {S.}~\bibnamefont {Agrestini}}, \bibinfo {author} {\bibfnamefont {Z.~X.}\ \bibnamefont {Shen}}, \bibinfo {author} {\bibfnamefont {E.~M.}\ \bibnamefont {Been}}, \bibinfo {author} {\bibfnamefont {B.}~\bibnamefont {Moritz}}, \bibinfo {author} {\bibfnamefont {T.~P.}\ \bibnamefont {Devereaux}}, \bibinfo {author} {\bibfnamefont {J.}~\bibnamefont {Zaanen}}, \bibinfo {author} {\bibfnamefont {H.~Y.}\ \bibnamefont {Hwang}}, \bibinfo {author} {\bibfnamefont {K.-J.}\ \bibnamefont
  {Zhou}},\ and\ \bibinfo {author} {\bibfnamefont {W.~S.}\ \bibnamefont {Lee}},\ }\bibfield  {title} {\bibinfo {title} {{Magnetic excitations in infinite-layer nickelates}},\ }\href {https://doi.org/10.1126/science.abd7726} {\bibfield  {journal} {\bibinfo  {journal} {Science}\ }\textbf {\bibinfo {volume} {373}},\ \bibinfo {pages} {213} (\bibinfo {year} {2021})}\BibitemShut {NoStop}%
\bibitem [{\citenamefont {{Worm}}\ \emph {et~al.}(2023)\citenamefont {{Worm}}, \citenamefont {{Wang}}, \citenamefont {{Kitatani}}, \citenamefont {{Bia{\l}o}}, \citenamefont {{Gao}}, \citenamefont {{Ren}}, \citenamefont {{Choi}}, \citenamefont {{Csontosov{\'a}}}, \citenamefont {{Zhou}}, \citenamefont {{Zhou}}, \citenamefont {{Zhu}}, \citenamefont {{Si}}, \citenamefont {{Chang}}, \citenamefont {{Tomczak}},\ and\ \citenamefont {{Held}}}]{Worm2024}%
  \BibitemOpen
  \bibfield  {author} {\bibinfo {author} {\bibfnamefont {P.}~\bibnamefont {{Worm}}}, \bibinfo {author} {\bibfnamefont {Q.}~\bibnamefont {{Wang}}}, \bibinfo {author} {\bibfnamefont {M.}~\bibnamefont {{Kitatani}}}, \bibinfo {author} {\bibfnamefont {I.}~\bibnamefont {{Bia{\l}o}}}, \bibinfo {author} {\bibfnamefont {Q.}~\bibnamefont {{Gao}}}, \bibinfo {author} {\bibfnamefont {X.}~\bibnamefont {{Ren}}}, \bibinfo {author} {\bibfnamefont {J.}~\bibnamefont {{Choi}}}, \bibinfo {author} {\bibfnamefont {D.}~\bibnamefont {{Csontosov{\'a}}}}, \bibinfo {author} {\bibfnamefont {K.-J.}\ \bibnamefont {{Zhou}}}, \bibinfo {author} {\bibfnamefont {X.}~\bibnamefont {{Zhou}}}, \bibinfo {author} {\bibfnamefont {Z.}~\bibnamefont {{Zhu}}}, \bibinfo {author} {\bibfnamefont {L.}~\bibnamefont {{Si}}}, \bibinfo {author} {\bibfnamefont {J.}~\bibnamefont {{Chang}}}, \bibinfo {author} {\bibfnamefont {J.~M.}\ \bibnamefont {{Tomczak}}},\ and\ \bibinfo {author} {\bibfnamefont {K.}~\bibnamefont {{Held}}},\ }\bibfield  {title} {\bibinfo {title}
  {{Spin fluctuations sufficient to mediate superconductivity in nickelates}},\ }\bibfield  {journal} {\bibinfo  {journal} {arXiv:2312.08260}\ }\href {https://doi.org/10.48550/arXiv.2312.08260} {10.48550/arXiv.2312.08260} (\bibinfo {year} {2023})\BibitemShut {NoStop}%
\bibitem [{\citenamefont {Adhikary}\ \emph {et~al.}(2020)\citenamefont {Adhikary}, \citenamefont {Bandyopadhyay}, \citenamefont {Das}, \citenamefont {Dasgupta},\ and\ \citenamefont {Saha-Dasgupta}}]{Adhikary2020}%
  \BibitemOpen
  \bibfield  {author} {\bibinfo {author} {\bibfnamefont {P.}~\bibnamefont {Adhikary}}, \bibinfo {author} {\bibfnamefont {S.}~\bibnamefont {Bandyopadhyay}}, \bibinfo {author} {\bibfnamefont {T.}~\bibnamefont {Das}}, \bibinfo {author} {\bibfnamefont {I.}~\bibnamefont {Dasgupta}},\ and\ \bibinfo {author} {\bibfnamefont {T.}~\bibnamefont {Saha-Dasgupta}},\ }\bibfield  {title} {\bibinfo {title} {{Orbital-selective superconductivity in a two-band model of infinite-layer nickelates}},\ }\href {https://doi.org/10.1103/PhysRevB.102.100501} {\bibfield  {journal} {\bibinfo  {journal} {Phys. Rev. B}\ }\textbf {\bibinfo {volume} {102}},\ \bibinfo {pages} {100501} (\bibinfo {year} {2020})}\BibitemShut {NoStop}%
\bibitem [{\citenamefont {Wang}\ \emph {et~al.}(2020{\natexlab{a}})\citenamefont {Wang}, \citenamefont {Kang}, \citenamefont {Miao},\ and\ \citenamefont {Kotliar}}]{Wang2020t}%
  \BibitemOpen
  \bibfield  {author} {\bibinfo {author} {\bibfnamefont {Y.}~\bibnamefont {Wang}}, \bibinfo {author} {\bibfnamefont {C.-J.}\ \bibnamefont {Kang}}, \bibinfo {author} {\bibfnamefont {H.}~\bibnamefont {Miao}},\ and\ \bibinfo {author} {\bibfnamefont {G.}~\bibnamefont {Kotliar}},\ }\bibfield  {title} {\bibinfo {title} {Hund's metal physics: From ${\mathrm{srnio}}_{2}$ to ${\mathrm{lanio}}_{2}$},\ }\href {https://doi.org/10.1103/PhysRevB.102.161118} {\bibfield  {journal} {\bibinfo  {journal} {Phys. Rev. B}\ }\textbf {\bibinfo {volume} {102}},\ \bibinfo {pages} {161118} (\bibinfo {year} {2020}{\natexlab{a}})}\BibitemShut {NoStop}%
\bibitem [{\citenamefont {Wang}\ \emph {et~al.}(2020{\natexlab{b}})\citenamefont {Wang}, \citenamefont {Zhang}, \citenamefont {Yang},\ and\ \citenamefont {Zhang}}]{WangZ2020}%
  \BibitemOpen
  \bibfield  {author} {\bibinfo {author} {\bibfnamefont {Z.}~\bibnamefont {Wang}}, \bibinfo {author} {\bibfnamefont {G.-M.}\ \bibnamefont {Zhang}}, \bibinfo {author} {\bibfnamefont {Y.-f.}\ \bibnamefont {Yang}},\ and\ \bibinfo {author} {\bibfnamefont {F.-C.}\ \bibnamefont {Zhang}},\ }\bibfield  {title} {\bibinfo {title} {Distinct pairing symmetries of superconductivity in infinite-layer nickelates},\ }\href {https://doi.org/10.1103/PhysRevB.102.220501} {\bibfield  {journal} {\bibinfo  {journal} {Phys. Rev. B}\ }\textbf {\bibinfo {volume} {102}},\ \bibinfo {pages} {220501} (\bibinfo {year} {2020}{\natexlab{b}})}\BibitemShut {NoStop}%
\bibitem [{\citenamefont {Kreisel}\ \emph {et~al.}(2022)\citenamefont {Kreisel}, \citenamefont {Andersen}, \citenamefont {R\o{}mer}, \citenamefont {Eremin},\ and\ \citenamefont {Lechermann}}]{Kreisel2022}%
  \BibitemOpen
  \bibfield  {author} {\bibinfo {author} {\bibfnamefont {A.}~\bibnamefont {Kreisel}}, \bibinfo {author} {\bibfnamefont {B.~M.}\ \bibnamefont {Andersen}}, \bibinfo {author} {\bibfnamefont {A.~T.}\ \bibnamefont {R\o{}mer}}, \bibinfo {author} {\bibfnamefont {I.~M.}\ \bibnamefont {Eremin}},\ and\ \bibinfo {author} {\bibfnamefont {F.}~\bibnamefont {Lechermann}},\ }\bibfield  {title} {\bibinfo {title} {Superconducting instabilities in strongly correlated infinite-layer nickelates},\ }\href {https://doi.org/10.1103/PhysRevLett.129.077002} {\bibfield  {journal} {\bibinfo  {journal} {Phys. Rev. Lett.}\ }\textbf {\bibinfo {volume} {129}},\ \bibinfo {pages} {077002} (\bibinfo {year} {2022})}\BibitemShut {NoStop}%
\bibitem [{\citenamefont {Lechermann}(2020{\natexlab{a}})}]{Lechermann2019}%
  \BibitemOpen
  \bibfield  {author} {\bibinfo {author} {\bibfnamefont {F.}~\bibnamefont {Lechermann}},\ }\bibfield  {title} {\bibinfo {title} {{Late transition metal oxides with infinite-layer structure: Nickelates versus cuprates}},\ }\href {https://doi.org/10.1103/PhysRevB.101.081110} {\bibfield  {journal} {\bibinfo  {journal} {Phys. Rev. B}\ }\textbf {\bibinfo {volume} {101}},\ \bibinfo {pages} {81110} (\bibinfo {year} {2020}{\natexlab{a}})}\BibitemShut {NoStop}%
\bibitem [{\citenamefont {Lechermann}(2020{\natexlab{b}})}]{Lechermann2020}%
  \BibitemOpen
  \bibfield  {author} {\bibinfo {author} {\bibfnamefont {F.}~\bibnamefont {Lechermann}},\ }\bibfield  {title} {\bibinfo {title} {{Multiorbital Processes Rule the {${\mathrm{Nd}}_{1\ensuremath{-}x}{\mathrm{Sr}}_{x}{\mathrm{NiO}}_{2}$} Normal State}},\ }\href {https://doi.org/10.1103/PhysRevX.10.041002} {\bibfield  {journal} {\bibinfo  {journal} {Phys. Rev. X}\ }\textbf {\bibinfo {volume} {10}},\ \bibinfo {pages} {41002} (\bibinfo {year} {2020}{\natexlab{b}})}\BibitemShut {NoStop}%
\bibitem [{\citenamefont {Lechermann}(2021)}]{Lechermann2021}%
  \BibitemOpen
  \bibfield  {author} {\bibinfo {author} {\bibfnamefont {F.}~\bibnamefont {Lechermann}},\ }\bibfield  {title} {\bibinfo {title} {Doping-dependent character and possible magnetic ordering of ndnio 2},\ }\href {https://doi.org/10.1103/PhysRevMaterials.5.044803} {\bibfield  {journal} {\bibinfo  {journal} {Physical Review Materials}\ }\textbf {\bibinfo {volume} {5}},\ \bibinfo {pages} {044803} (\bibinfo {year} {2021})}\BibitemShut {NoStop}%
\bibitem [{\citenamefont {Petocchi}\ \emph {et~al.}(2020)\citenamefont {Petocchi}, \citenamefont {Christiansson}, \citenamefont {Nilsson}, \citenamefont {Aryasetiawan},\ and\ \citenamefont {Werner}}]{Petocchi2020}%
  \BibitemOpen
  \bibfield  {author} {\bibinfo {author} {\bibfnamefont {F.}~\bibnamefont {Petocchi}}, \bibinfo {author} {\bibfnamefont {V.}~\bibnamefont {Christiansson}}, \bibinfo {author} {\bibfnamefont {F.}~\bibnamefont {Nilsson}}, \bibinfo {author} {\bibfnamefont {F.}~\bibnamefont {Aryasetiawan}},\ and\ \bibinfo {author} {\bibfnamefont {P.}~\bibnamefont {Werner}},\ }\bibfield  {title} {\bibinfo {title} {{Normal State of ${\mathrm{Nd}}_{1\ensuremath{-}x}{\mathrm{Sr}}_{x}{\mathrm{NiO}}_{2}$ from Self-Consistent $GW+\mathrm{EDMFT}$}},\ }\href {https://doi.org/10.1103/PhysRevX.10.041047} {\bibfield  {journal} {\bibinfo  {journal} {Phys. Rev. X}\ }\textbf {\bibinfo {volume} {10}},\ \bibinfo {pages} {41047} (\bibinfo {year} {2020})}\BibitemShut {NoStop}%
\bibitem [{\citenamefont {Wan}\ \emph {et~al.}(2021)\citenamefont {Wan}, \citenamefont {Ivanov}, \citenamefont {Resta}, \citenamefont {Leonov},\ and\ \citenamefont {Savrasov}}]{Wan2021}%
  \BibitemOpen
  \bibfield  {author} {\bibinfo {author} {\bibfnamefont {X.}~\bibnamefont {Wan}}, \bibinfo {author} {\bibfnamefont {V.}~\bibnamefont {Ivanov}}, \bibinfo {author} {\bibfnamefont {G.}~\bibnamefont {Resta}}, \bibinfo {author} {\bibfnamefont {I.}~\bibnamefont {Leonov}},\ and\ \bibinfo {author} {\bibfnamefont {S.~Y.}\ \bibnamefont {Savrasov}},\ }\bibfield  {title} {\bibinfo {title} {Exchange interactions and sensitivity of the ni two-hole spin state to hund's coupling in doped ${\mathrm{ndnio}}_{2}$},\ }\href {https://doi.org/10.1103/PhysRevB.103.075123} {\bibfield  {journal} {\bibinfo  {journal} {Phys. Rev. B}\ }\textbf {\bibinfo {volume} {103}},\ \bibinfo {pages} {075123} (\bibinfo {year} {2021})}\BibitemShut {NoStop}%
\bibitem [{\citenamefont {Choi}\ \emph {et~al.}(2020)\citenamefont {Choi}, \citenamefont {Pickett},\ and\ \citenamefont {Lee}}]{MiYoung2020}%
  \BibitemOpen
  \bibfield  {author} {\bibinfo {author} {\bibfnamefont {M.-Y.}\ \bibnamefont {Choi}}, \bibinfo {author} {\bibfnamefont {W.~E.}\ \bibnamefont {Pickett}},\ and\ \bibinfo {author} {\bibfnamefont {K.-W.}\ \bibnamefont {Lee}},\ }\bibfield  {title} {\bibinfo {title} {Fluctuation-frustrated flat band instabilities in ${\mathrm{ndnio}}_{2}$},\ }\href {https://doi.org/10.1103/PhysRevResearch.2.033445} {\bibfield  {journal} {\bibinfo  {journal} {Phys. Rev. Research}\ }\textbf {\bibinfo {volume} {2}},\ \bibinfo {pages} {033445} (\bibinfo {year} {2020})}\BibitemShut {NoStop}%
\bibitem [{\citenamefont {{Sun}}\ \emph {et~al.}(2024)\citenamefont {{Sun}}, \citenamefont {{Jiang}}, \citenamefont {{Xia}}, \citenamefont {{Hao}}, \citenamefont {{Li}}, \citenamefont {{Yan}}, \citenamefont {{Wang}}, \citenamefont {{Liu}}, \citenamefont {{Ding}}, \citenamefont {{Liu}}, \citenamefont {{Liu}}, \citenamefont {{Liu}}, \citenamefont {{Chen}}, \citenamefont {{Shen}},\ and\ \citenamefont {{Nie}}}]{Sun2024}%
  \BibitemOpen
  \bibfield  {author} {\bibinfo {author} {\bibfnamefont {W.}~\bibnamefont {{Sun}}}, \bibinfo {author} {\bibfnamefont {Z.}~\bibnamefont {{Jiang}}}, \bibinfo {author} {\bibfnamefont {C.}~\bibnamefont {{Xia}}}, \bibinfo {author} {\bibfnamefont {B.}~\bibnamefont {{Hao}}}, \bibinfo {author} {\bibfnamefont {Y.}~\bibnamefont {{Li}}}, \bibinfo {author} {\bibfnamefont {S.}~\bibnamefont {{Yan}}}, \bibinfo {author} {\bibfnamefont {M.}~\bibnamefont {{Wang}}}, \bibinfo {author} {\bibfnamefont {H.}~\bibnamefont {{Liu}}}, \bibinfo {author} {\bibfnamefont {J.}~\bibnamefont {{Ding}}}, \bibinfo {author} {\bibfnamefont {J.}~\bibnamefont {{Liu}}}, \bibinfo {author} {\bibfnamefont {Z.}~\bibnamefont {{Liu}}}, \bibinfo {author} {\bibfnamefont {J.}~\bibnamefont {{Liu}}}, \bibinfo {author} {\bibfnamefont {H.}~\bibnamefont {{Chen}}}, \bibinfo {author} {\bibfnamefont {D.}~\bibnamefont {{Shen}}},\ and\ \bibinfo {author} {\bibfnamefont {Y.}~\bibnamefont {{Nie}}},\ }\bibfield  {title} {\bibinfo {title} {{Electronic Structure of
  Superconducting Infinite-Layer Lanthanum Nickelates}},\ }\bibfield  {journal} {\bibinfo  {journal} {arXiv:2403.07344}\ }\href {https://doi.org/10.48550/arXiv.2403.07344} {10.48550/arXiv.2403.07344} (\bibinfo {year} {2024})\BibitemShut {NoStop}%
\bibitem [{\citenamefont {Ding}\ \emph {et~al.}(2024)\citenamefont {Ding}, \citenamefont {Fan}, \citenamefont {Wang}, \citenamefont {Li}, \citenamefont {An}, \citenamefont {Ye}, \citenamefont {Tang}, \citenamefont {Lei}, \citenamefont {Sun}, \citenamefont {Guo}, \citenamefont {Chen}, \citenamefont {Sangphet}, \citenamefont {Wang}, \citenamefont {Xu}, \citenamefont {Peng},\ and\ \citenamefont {Feng}}]{Ding2024}%
  \BibitemOpen
  \bibfield  {author} {\bibinfo {author} {\bibfnamefont {X.}~\bibnamefont {Ding}}, \bibinfo {author} {\bibfnamefont {Y.}~\bibnamefont {Fan}}, \bibinfo {author} {\bibfnamefont {X.}~\bibnamefont {Wang}}, \bibinfo {author} {\bibfnamefont {C.}~\bibnamefont {Li}}, \bibinfo {author} {\bibfnamefont {Z.}~\bibnamefont {An}}, \bibinfo {author} {\bibfnamefont {J.}~\bibnamefont {Ye}}, \bibinfo {author} {\bibfnamefont {S.}~\bibnamefont {Tang}}, \bibinfo {author} {\bibfnamefont {M.}~\bibnamefont {Lei}}, \bibinfo {author} {\bibfnamefont {X.}~\bibnamefont {Sun}}, \bibinfo {author} {\bibfnamefont {N.}~\bibnamefont {Guo}}, \bibinfo {author} {\bibfnamefont {Z.}~\bibnamefont {Chen}}, \bibinfo {author} {\bibfnamefont {S.}~\bibnamefont {Sangphet}}, \bibinfo {author} {\bibfnamefont {Y.}~\bibnamefont {Wang}}, \bibinfo {author} {\bibfnamefont {H.}~\bibnamefont {Xu}}, \bibinfo {author} {\bibfnamefont {R.}~\bibnamefont {Peng}},\ and\ \bibinfo {author} {\bibfnamefont {D.}~\bibnamefont {Feng}},\ }\bibfield  {title} {\bibinfo {title}
  {{Cuprate-like electronic structures in infinite-layer nickelates with substantial hole dopings}},\ }\href {https://doi.org/10.1093/nsr/nwae194} {\bibfield  {journal} {\bibinfo  {journal} {National Science Review}\ ,\ \bibinfo {pages} {nwae194}} (\bibinfo {year} {2024})}\BibitemShut {NoStop}%
\bibitem [{Note2()}]{Note2}%
  \BibitemOpen
  \bibinfo {note} {As in earlier photoemission spectroscopy (PES) experiments \cite {Chen2022}}\BibitemShut {NoStop}%
\bibitem [{\citenamefont {Blaha}\ \emph {et~al.}(2019)\citenamefont {Blaha}, \citenamefont {Schwarz}, \citenamefont {Madsen}, \citenamefont {Kvasnicka}, \citenamefont {Luitz}, \citenamefont {Laskowsk}, \citenamefont {Tran}, \citenamefont {Marks},\ and\ \citenamefont {Marks}}]{blaha2001wien2k}%
  \BibitemOpen
  \bibfield  {author} {\bibinfo {author} {\bibfnamefont {P.}~\bibnamefont {Blaha}}, \bibinfo {author} {\bibfnamefont {K.}~\bibnamefont {Schwarz}}, \bibinfo {author} {\bibfnamefont {G.}~\bibnamefont {Madsen}}, \bibinfo {author} {\bibfnamefont {D.}~\bibnamefont {Kvasnicka}}, \bibinfo {author} {\bibfnamefont {J.}~\bibnamefont {Luitz}}, \bibinfo {author} {\bibfnamefont {R.}~\bibnamefont {Laskowsk}}, \bibinfo {author} {\bibfnamefont {F.}~\bibnamefont {Tran}}, \bibinfo {author} {\bibfnamefont {L.}~\bibnamefont {Marks}},\ and\ \bibinfo {author} {\bibfnamefont {L.}~\bibnamefont {Marks}},\ }\href@noop {} {\bibinfo {title} {Wien2k: An augmented plane wave plus local orbitals program for calculating crystal properties}} (\bibinfo {year} {2019})\BibitemShut {NoStop}%
\bibitem [{\citenamefont {Perdew}\ \emph {et~al.}(2008)\citenamefont {Perdew}, \citenamefont {Ruzsinszky}, \citenamefont {Csonka}, \citenamefont {Vydrov}, \citenamefont {Scuseria}, \citenamefont {Constantin}, \citenamefont {Zhou},\ and\ \citenamefont {Burke}}]{PhysRevLett.100.136406}%
  \BibitemOpen
  \bibfield  {author} {\bibinfo {author} {\bibfnamefont {J.~P.}\ \bibnamefont {Perdew}}, \bibinfo {author} {\bibfnamefont {A.}~\bibnamefont {Ruzsinszky}}, \bibinfo {author} {\bibfnamefont {G.~I.}\ \bibnamefont {Csonka}}, \bibinfo {author} {\bibfnamefont {O.~A.}\ \bibnamefont {Vydrov}}, \bibinfo {author} {\bibfnamefont {G.~E.}\ \bibnamefont {Scuseria}}, \bibinfo {author} {\bibfnamefont {L.~A.}\ \bibnamefont {Constantin}}, \bibinfo {author} {\bibfnamefont {X.}~\bibnamefont {Zhou}},\ and\ \bibinfo {author} {\bibfnamefont {K.}~\bibnamefont {Burke}},\ }\bibfield  {title} {\bibinfo {title} {Restoring the density-gradient expansion for exchange in solids and surfaces},\ }\href {https://doi.org/10.1103/PhysRevLett.100.136406} {\bibfield  {journal} {\bibinfo  {journal} {Phys. Rev. Lett.}\ }\textbf {\bibinfo {volume} {100}},\ \bibinfo {pages} {136406} (\bibinfo {year} {2008})}\BibitemShut {NoStop}%
\bibitem [{\citenamefont {Kune{\v{s}}}\ \emph {et~al.}(2010)\citenamefont {Kune{\v{s}}}, \citenamefont {Arita}, \citenamefont {Wissgott}, \citenamefont {Toschi}, \citenamefont {Ikeda},\ and\ \citenamefont {Held}}]{Kunes2010a}%
  \BibitemOpen
  \bibfield  {author} {\bibinfo {author} {\bibfnamefont {J.}~\bibnamefont {Kune{\v{s}}}}, \bibinfo {author} {\bibfnamefont {R.}~\bibnamefont {Arita}}, \bibinfo {author} {\bibfnamefont {P.}~\bibnamefont {Wissgott}}, \bibinfo {author} {\bibfnamefont {A.}~\bibnamefont {Toschi}}, \bibinfo {author} {\bibfnamefont {H.}~\bibnamefont {Ikeda}},\ and\ \bibinfo {author} {\bibfnamefont {K.}~\bibnamefont {Held}},\ }\bibfield  {title} {\bibinfo {title} {{Wien2wannier: From linearized augmented plane waves to maximally localized Wannier functions}},\ }\href {https://doi.org/https://doi.org/10.1016/j.cpc.2010.08.005} {\bibfield  {journal} {\bibinfo  {journal} {Comp. Phys. Comm.}\ }\textbf {\bibinfo {volume} {181}},\ \bibinfo {pages} {1888} (\bibinfo {year} {2010})}\BibitemShut {NoStop}%
\bibitem [{\citenamefont {Marzari}\ \emph {et~al.}(2012)\citenamefont {Marzari}, \citenamefont {Mostofi}, \citenamefont {Yates}, \citenamefont {Souza},\ and\ \citenamefont {Vanderbilt}}]{RevModPhys.84.1419}%
  \BibitemOpen
  \bibfield  {author} {\bibinfo {author} {\bibfnamefont {N.}~\bibnamefont {Marzari}}, \bibinfo {author} {\bibfnamefont {A.~A.}\ \bibnamefont {Mostofi}}, \bibinfo {author} {\bibfnamefont {J.~R.}\ \bibnamefont {Yates}}, \bibinfo {author} {\bibfnamefont {I.}~\bibnamefont {Souza}},\ and\ \bibinfo {author} {\bibfnamefont {D.}~\bibnamefont {Vanderbilt}},\ }\bibfield  {title} {\bibinfo {title} {Maximally localized wannier functions: Theory and applications},\ }\href {https://doi.org/10.1103/RevModPhys.84.1419} {\bibfield  {journal} {\bibinfo  {journal} {Rev. Mod. Phys.}\ }\textbf {\bibinfo {volume} {84}},\ \bibinfo {pages} {1419} (\bibinfo {year} {2012})}\BibitemShut {NoStop}%
\bibitem [{\citenamefont {Mostofi}\ \emph {et~al.}(2008)\citenamefont {Mostofi}, \citenamefont {Yates}, \citenamefont {Lee}, \citenamefont {Souza}, \citenamefont {Vanderbilt},\ and\ \citenamefont {Marzari}}]{mostofi2008wannier90}%
  \BibitemOpen
  \bibfield  {author} {\bibinfo {author} {\bibfnamefont {A.~A.}\ \bibnamefont {Mostofi}}, \bibinfo {author} {\bibfnamefont {J.~R.}\ \bibnamefont {Yates}}, \bibinfo {author} {\bibfnamefont {Y.-S.}\ \bibnamefont {Lee}}, \bibinfo {author} {\bibfnamefont {I.}~\bibnamefont {Souza}}, \bibinfo {author} {\bibfnamefont {D.}~\bibnamefont {Vanderbilt}},\ and\ \bibinfo {author} {\bibfnamefont {N.}~\bibnamefont {Marzari}},\ }\bibfield  {title} {\bibinfo {title} {wannier90: A tool for obtaining maximally-localised wannier functions},\ }\href {https://doi.org/https://doi.org/10.1016/j.cpc.2007.11.016} {\bibfield  {journal} {\bibinfo  {journal} {Computer Physics Communications}\ }\textbf {\bibinfo {volume} {178}},\ \bibinfo {pages} {685} (\bibinfo {year} {2008})}\BibitemShut {NoStop}%
\bibitem [{\citenamefont {Pizzi}\ \emph {et~al.}(2020)\citenamefont {Pizzi}, \citenamefont {Vitale}, \citenamefont {Arita}, \citenamefont {Blügel}, \citenamefont {Freimuth}, \citenamefont {G{\'{e}}ranton}, \citenamefont {Gibertini}, \citenamefont {Gresch}, \citenamefont {Johnson}, \citenamefont {Koretsune}, \citenamefont {Iba{\~{n}}ez-Azpiroz}, \citenamefont {Lee}, \citenamefont {Lihm}, \citenamefont {Marchand}, \citenamefont {Marrazzo}, \citenamefont {Mokrousov}, \citenamefont {Mustafa}, \citenamefont {Nohara}, \citenamefont {Nomura}, \citenamefont {Paulatto}, \citenamefont {Ponc{\'{e}}}, \citenamefont {Ponweiser}, \citenamefont {Qiao}, \citenamefont {Thöle}, \citenamefont {Tsirkin}, \citenamefont {Wierzbowska}, \citenamefont {Marzari}, \citenamefont {Vanderbilt}, \citenamefont {Souza}, \citenamefont {Mostofi},\ and\ \citenamefont {Yates}}]{Pizzi2020}%
  \BibitemOpen
  \bibfield  {author} {\bibinfo {author} {\bibfnamefont {G.}~\bibnamefont {Pizzi}}, \bibinfo {author} {\bibfnamefont {V.}~\bibnamefont {Vitale}}, \bibinfo {author} {\bibfnamefont {R.}~\bibnamefont {Arita}}, \bibinfo {author} {\bibfnamefont {S.}~\bibnamefont {Blügel}}, \bibinfo {author} {\bibfnamefont {F.}~\bibnamefont {Freimuth}}, \bibinfo {author} {\bibfnamefont {G.}~\bibnamefont {G{\'{e}}ranton}}, \bibinfo {author} {\bibfnamefont {M.}~\bibnamefont {Gibertini}}, \bibinfo {author} {\bibfnamefont {D.}~\bibnamefont {Gresch}}, \bibinfo {author} {\bibfnamefont {C.}~\bibnamefont {Johnson}}, \bibinfo {author} {\bibfnamefont {T.}~\bibnamefont {Koretsune}}, \bibinfo {author} {\bibfnamefont {J.}~\bibnamefont {Iba{\~{n}}ez-Azpiroz}}, \bibinfo {author} {\bibfnamefont {H.}~\bibnamefont {Lee}}, \bibinfo {author} {\bibfnamefont {J.-M.}\ \bibnamefont {Lihm}}, \bibinfo {author} {\bibfnamefont {D.}~\bibnamefont {Marchand}}, \bibinfo {author} {\bibfnamefont {A.}~\bibnamefont {Marrazzo}}, \bibinfo {author} {\bibfnamefont
  {Y.}~\bibnamefont {Mokrousov}}, \bibinfo {author} {\bibfnamefont {J.~I.}\ \bibnamefont {Mustafa}}, \bibinfo {author} {\bibfnamefont {Y.}~\bibnamefont {Nohara}}, \bibinfo {author} {\bibfnamefont {Y.}~\bibnamefont {Nomura}}, \bibinfo {author} {\bibfnamefont {L.}~\bibnamefont {Paulatto}}, \bibinfo {author} {\bibfnamefont {S.}~\bibnamefont {Ponc{\'{e}}}}, \bibinfo {author} {\bibfnamefont {T.}~\bibnamefont {Ponweiser}}, \bibinfo {author} {\bibfnamefont {J.}~\bibnamefont {Qiao}}, \bibinfo {author} {\bibfnamefont {F.}~\bibnamefont {Thöle}}, \bibinfo {author} {\bibfnamefont {S.~S.}\ \bibnamefont {Tsirkin}}, \bibinfo {author} {\bibfnamefont {M.}~\bibnamefont {Wierzbowska}}, \bibinfo {author} {\bibfnamefont {N.}~\bibnamefont {Marzari}}, \bibinfo {author} {\bibfnamefont {D.}~\bibnamefont {Vanderbilt}}, \bibinfo {author} {\bibfnamefont {I.}~\bibnamefont {Souza}}, \bibinfo {author} {\bibfnamefont {A.~A.}\ \bibnamefont {Mostofi}},\ and\ \bibinfo {author} {\bibfnamefont {J.~R.}\ \bibnamefont {Yates}},\ }\bibfield
  {title} {\bibinfo {title} {Wannier90 as a community code: new features and applications},\ }\href {https://doi.org/10.1088/1361-648x/ab51ff} {\bibfield  {journal} {\bibinfo  {journal} {Journal of Physics: Condensed Matter}\ }\textbf {\bibinfo {volume} {32}},\ \bibinfo {pages} {165902} (\bibinfo {year} {2020})}\BibitemShut {NoStop}%
\bibitem [{\citenamefont {Miyake}\ and\ \citenamefont {Aryasetiawan}(2008)}]{PhysRevB.77.085122}%
  \BibitemOpen
  \bibfield  {author} {\bibinfo {author} {\bibfnamefont {T.}~\bibnamefont {Miyake}}\ and\ \bibinfo {author} {\bibfnamefont {F.}~\bibnamefont {Aryasetiawan}},\ }\bibfield  {title} {\bibinfo {title} {{Screened Coulomb interaction in the maximally localized Wannier basis}},\ }\href {https://doi.org/10.1103/PhysRevB.77.085122} {\bibfield  {journal} {\bibinfo  {journal} {Phys. Rev. B}\ }\textbf {\bibinfo {volume} {77}},\ \bibinfo {pages} {85122} (\bibinfo {year} {2008})}\BibitemShut {NoStop}%
\bibitem [{\citenamefont {Anisimov}\ \emph {et~al.}(1991)\citenamefont {Anisimov}, \citenamefont {Zaanen},\ and\ \citenamefont {Andersen}}]{Anisimov1991}%
  \BibitemOpen
  \bibfield  {author} {\bibinfo {author} {\bibfnamefont {V.~I.}\ \bibnamefont {Anisimov}}, \bibinfo {author} {\bibfnamefont {J.}~\bibnamefont {Zaanen}},\ and\ \bibinfo {author} {\bibfnamefont {O.~K.}\ \bibnamefont {Andersen}},\ }\bibfield  {title} {\bibinfo {title} {{Band theory and Mott insulators: Hubbard U instead of Stoner I}},\ }\href {https://doi.org/10.1103/PhysRevB.44.943} {\bibfield  {journal} {\bibinfo  {journal} {Phys. Rev. B}\ }\textbf {\bibinfo {volume} {44}},\ \bibinfo {pages} {943} (\bibinfo {year} {1991})}\BibitemShut {NoStop}%
\bibitem [{\citenamefont {Kotliar}\ \emph {et~al.}(2006)\citenamefont {Kotliar}, \citenamefont {Savrasov}, \citenamefont {Haule}, \citenamefont {Oudovenko}, \citenamefont {Parcollet},\ and\ \citenamefont {Marianetti}}]{Kotliar2006}%
  \BibitemOpen
  \bibfield  {author} {\bibinfo {author} {\bibfnamefont {G.}~\bibnamefont {Kotliar}}, \bibinfo {author} {\bibfnamefont {S.~Y.}\ \bibnamefont {Savrasov}}, \bibinfo {author} {\bibfnamefont {K.}~\bibnamefont {Haule}}, \bibinfo {author} {\bibfnamefont {V.~S.}\ \bibnamefont {Oudovenko}}, \bibinfo {author} {\bibfnamefont {O.}~\bibnamefont {Parcollet}},\ and\ \bibinfo {author} {\bibfnamefont {C.~A.}\ \bibnamefont {Marianetti}},\ }\bibfield  {title} {\bibinfo {title} {{Electronic structure calculations with dynamical mean-field theory}},\ }\href {https://doi.org/10.1103/RevModPhys.78.865} {\bibfield  {journal} {\bibinfo  {journal} {Rev. Mod. Phys.}\ }\textbf {\bibinfo {volume} {78}},\ \bibinfo {pages} {865} (\bibinfo {year} {2006})}\BibitemShut {NoStop}%
\bibitem [{\citenamefont {Gull}\ \emph {et~al.}(2011)\citenamefont {Gull}, \citenamefont {Millis}, \citenamefont {Lichtenstein}, \citenamefont {Rubtsov}, \citenamefont {Troyer},\ and\ \citenamefont {Werner}}]{Gull2011a}%
  \BibitemOpen
  \bibfield  {author} {\bibinfo {author} {\bibfnamefont {E.}~\bibnamefont {Gull}}, \bibinfo {author} {\bibfnamefont {A.~J.}\ \bibnamefont {Millis}}, \bibinfo {author} {\bibfnamefont {A.~I.}\ \bibnamefont {Lichtenstein}}, \bibinfo {author} {\bibfnamefont {A.~N.}\ \bibnamefont {Rubtsov}}, \bibinfo {author} {\bibfnamefont {M.}~\bibnamefont {Troyer}},\ and\ \bibinfo {author} {\bibfnamefont {P.}~\bibnamefont {Werner}},\ }\bibfield  {title} {\bibinfo {title} {{Continuous-time Monte Carlo methods for quantum impurity models}},\ }\href {https://doi.org/10.1103/RevModPhys.83.349} {\bibfield  {journal} {\bibinfo  {journal} {Rev. Mod. Phys.}\ }\textbf {\bibinfo {volume} {83}},\ \bibinfo {pages} {349} (\bibinfo {year} {2011})}\BibitemShut {NoStop}%
\bibitem [{\citenamefont {Wallerberger}\ \emph {et~al.}(2019)\citenamefont {Wallerberger}, \citenamefont {Hausoel}, \citenamefont {Gunacker}, \citenamefont {Parragh}, \citenamefont {Goth}, \citenamefont {Held},\ and\ \citenamefont {Sangiovanni}}]{w2dynamics2018}%
  \BibitemOpen
  \bibfield  {author} {\bibinfo {author} {\bibfnamefont {M.}~\bibnamefont {Wallerberger}}, \bibinfo {author} {\bibfnamefont {A.}~\bibnamefont {Hausoel}}, \bibinfo {author} {\bibfnamefont {A.}~\bibnamefont {Gunacker}, \bibfnamefont {Patrik fand~Kowalski}}, \bibinfo {author} {\bibfnamefont {N.}~\bibnamefont {Parragh}}, \bibinfo {author} {\bibfnamefont {F.}~\bibnamefont {Goth}}, \bibinfo {author} {\bibfnamefont {K.}~\bibnamefont {Held}},\ and\ \bibinfo {author} {\bibfnamefont {G.}~\bibnamefont {Sangiovanni}},\ }\bibfield  {title} {\bibinfo {title} {{w2dynamics: Local one- and two-particle quantities from dynamical mean field theory}},\ }\href {https://doi.org/https://doi.org/10.1016/j.cpc.2018.09.007} {\bibfield  {journal} {\bibinfo  {journal} {Comp. Phys. Comm.}\ }\textbf {\bibinfo {volume} {235}},\ \bibinfo {pages} {388} (\bibinfo {year} {2019})}\BibitemShut {NoStop}%
\bibitem [{\citenamefont {Gubernatis}\ \emph {et~al.}(1991)\citenamefont {Gubernatis}, \citenamefont {Jarrell}, \citenamefont {Silver},\ and\ \citenamefont {Sivia}}]{PhysRevB.44.6011}%
  \BibitemOpen
  \bibfield  {author} {\bibinfo {author} {\bibfnamefont {J.~E.}\ \bibnamefont {Gubernatis}}, \bibinfo {author} {\bibfnamefont {M.}~\bibnamefont {Jarrell}}, \bibinfo {author} {\bibfnamefont {R.~N.}\ \bibnamefont {Silver}},\ and\ \bibinfo {author} {\bibfnamefont {D.~S.}\ \bibnamefont {Sivia}},\ }\bibfield  {title} {\bibinfo {title} {{Quantum Monte Carlo simulations and maximum entropy: Dynamics from imaginary-time data}},\ }\href {https://doi.org/10.1103/PhysRevB.44.6011} {\bibfield  {journal} {\bibinfo  {journal} {Phys. Rev. B}\ }\textbf {\bibinfo {volume} {44}},\ \bibinfo {pages} {6011} (\bibinfo {year} {1991})}\BibitemShut {NoStop}%
\bibitem [{\citenamefont {Kaufmann}\ and\ \citenamefont {Held}(2023)}]{Kaufmann2021}%
  \BibitemOpen
  \bibfield  {author} {\bibinfo {author} {\bibfnamefont {J.}~\bibnamefont {Kaufmann}}\ and\ \bibinfo {author} {\bibfnamefont {K.}~\bibnamefont {Held}},\ }\bibfield  {title} {\bibinfo {title} {ana\_cont: Python package for analytic continuation},\ }\href {https://doi.org/https://doi.org/10.1016/j.cpc.2022.108519} {\bibfield  {journal} {\bibinfo  {journal} {Comp. Phys. Comm.}\ }\textbf {\bibinfo {volume} {282}},\ \bibinfo {pages} {108519} (\bibinfo {year} {2023})}\BibitemShut {NoStop}%
\bibitem [{Note3()}]{Note3}%
  \BibitemOpen
  \bibinfo {note} {See Fig.~1 of \cite {Sun2024} and Fig.~1 of \cite {Ding2024}}\BibitemShut {NoStop}%
\bibitem [{Note4()}]{Note4}%
  \BibitemOpen
  \bibinfo {note} {See e.g.~Fig.~11 of \cite {Lechermann2020} and Fig.~4(f) of \cite {Petocchi2020}}\BibitemShut {NoStop}%
\bibitem [{Note5()}]{Note5}%
  \BibitemOpen
  \bibinfo {note} {See Fig.~3(b) of \cite {Lechermann2020} and Fig.~4(a) of \cite {Petocchi2020}}\BibitemShut {NoStop}%
\bibitem [{\citenamefont {Sun}\ and\ \citenamefont {Kotliar}(2004)}]{Sun2004}%
  \BibitemOpen
  \bibfield  {author} {\bibinfo {author} {\bibfnamefont {P.}~\bibnamefont {Sun}}\ and\ \bibinfo {author} {\bibfnamefont {G.}~\bibnamefont {Kotliar}},\ }\bibfield  {title} {\bibinfo {title} {Many-body approximation scheme beyond gw},\ }\href {https://doi.org/10.1103/PhysRevLett.92.196402} {\bibfield  {journal} {\bibinfo  {journal} {Phys. Rev. Lett.}\ }\textbf {\bibinfo {volume} {92}},\ \bibinfo {pages} {196402} (\bibinfo {year} {2004})}\BibitemShut {NoStop}%
\bibitem [{\citenamefont {{First version of \protect \cite{Ding2024}}}()}]{DingV1}%
  \BibitemOpen
  \bibfield  {author} {\bibinfo {author} {\bibnamefont {{First version of \protect \cite{Ding2024}}}},\ }\href {https://arxiv.org/pdf/2403.07448v1} {\bibinfo  {journal} {arXiv:2403.07448v1}\ }\BibitemShut {NoStop}%
\bibitem [{\citenamefont {Si}\ \emph {et~al.}(2022)\citenamefont {Si}, \citenamefont {Worm},\ and\ \citenamefont {Held}}]{Si2022a}%
  \BibitemOpen
\bibfield  {journal} {  }\bibfield  {author} {\bibinfo {author} {\bibfnamefont {L.}~\bibnamefont {Si}}, \bibinfo {author} {\bibfnamefont {P.}~\bibnamefont {Worm}},\ and\ \bibinfo {author} {\bibfnamefont {K.}~\bibnamefont {Held}},\ }\bibfield  {title} {\bibinfo {title} {Fingerprints of topotactic hydrogen in nickelate superconductors},\ }\bibfield  {journal} {\bibinfo  {journal} {Crystals}\ }\textbf {\bibinfo {volume} {12}},\ \href {https://doi.org/10.3390/cryst12050656} {10.3390/cryst12050656} (\bibinfo {year} {2022})\BibitemShut {NoStop}%
\bibitem [{\citenamefont {Si}\ \emph {et~al.}(2023)\citenamefont {Si}, \citenamefont {Worm}, \citenamefont {Chen},\ and\ \citenamefont {Held}}]{PhysRevB.107.165116}%
  \BibitemOpen
  \bibfield  {author} {\bibinfo {author} {\bibfnamefont {L.}~\bibnamefont {Si}}, \bibinfo {author} {\bibfnamefont {P.}~\bibnamefont {Worm}}, \bibinfo {author} {\bibfnamefont {D.}~\bibnamefont {Chen}},\ and\ \bibinfo {author} {\bibfnamefont {K.}~\bibnamefont {Held}},\ }\bibfield  {title} {\bibinfo {title} {Topotactic hydrogen forms chains in $ab{\mathrm{o}}_{2}$ nickelate superconductors},\ }\href {https://doi.org/10.1103/PhysRevB.107.165116} {\bibfield  {journal} {\bibinfo  {journal} {Phys. Rev. B}\ }\textbf {\bibinfo {volume} {107}},\ \bibinfo {pages} {165116} (\bibinfo {year} {2023})}\BibitemShut {NoStop}%
\bibitem [{\citenamefont {Di~Cataldo}\ \emph {et~al.}(2023)\citenamefont {Di~Cataldo}, \citenamefont {Worm}, \citenamefont {Si},\ and\ \citenamefont {Held}}]{DiCataldo2023}%
  \BibitemOpen
  \bibfield  {author} {\bibinfo {author} {\bibfnamefont {S.}~\bibnamefont {Di~Cataldo}}, \bibinfo {author} {\bibfnamefont {P.}~\bibnamefont {Worm}}, \bibinfo {author} {\bibfnamefont {L.}~\bibnamefont {Si}},\ and\ \bibinfo {author} {\bibfnamefont {K.}~\bibnamefont {Held}},\ }\bibfield  {title} {\bibinfo {title} {Absence of electron-phonon-mediated superconductivity in hydrogen-intercalated nickelates},\ }\href {https://doi.org/10.1103/PhysRevB.108.174512} {\bibfield  {journal} {\bibinfo  {journal} {Phys. Rev. B}\ }\textbf {\bibinfo {volume} {108}},\ \bibinfo {pages} {174512} (\bibinfo {year} {2023})}\BibitemShut {NoStop}%
\bibitem [{\citenamefont {{Di Cataldo}}\ \emph {et~al.}(2023)\citenamefont {{Di Cataldo}}, \citenamefont {{Worm}}, \citenamefont {{Tomczak}}, \citenamefont {{Si}},\ and\ \citenamefont {{Held}}}]{DiCataldo2023b}%
  \BibitemOpen
  \bibfield  {author} {\bibinfo {author} {\bibfnamefont {S.}~\bibnamefont {{Di Cataldo}}}, \bibinfo {author} {\bibfnamefont {P.}~\bibnamefont {{Worm}}}, \bibinfo {author} {\bibfnamefont {J.}~\bibnamefont {{Tomczak}}}, \bibinfo {author} {\bibfnamefont {L.}~\bibnamefont {{Si}}},\ and\ \bibinfo {author} {\bibfnamefont {K.}~\bibnamefont {{Held}}},\ }\bibfield  {title} {\bibinfo {title} {{Unconventional superconductivity without doping: infinite-layer nickelates under pressure}},\ }\href {https://doi.org/10.48550/arXiv:2311.06195} {\bibfield  {journal} {\bibinfo  {journal} {arXiv:2311.06195}\ } (\bibinfo {year} {2023})}\BibitemShut {NoStop}%
\bibitem [{\citenamefont {{Krsnik}}\ and\ \citenamefont {{Held}}(2024)}]{Krsnik2024}%
  \BibitemOpen
  \bibfield  {author} {\bibinfo {author} {\bibfnamefont {J.}~\bibnamefont {{Krsnik}}}\ and\ \bibinfo {author} {\bibfnamefont {K.}~\bibnamefont {{Held}}},\ }\bibfield  {title} {\bibinfo {title} {{Waterfalls: umbilical cords at the birth of Hubbard bands}},\ }\bibfield  {journal} {\bibinfo  {journal} {arXiv:2408.12884}\ }\href {https://doi.org/10.48550/arXiv.2408.12884} {10.48550/arXiv.2408.12884} (\bibinfo {year} {2024})\BibitemShut {NoStop}%
\bibitem [{\citenamefont {Macridin}\ \emph {et~al.}(2007)\citenamefont {Macridin}, \citenamefont {Jarrell}, \citenamefont {Maier},\ and\ \citenamefont {Scalapino}}]{Macridin2007}%
  \BibitemOpen
  \bibfield  {author} {\bibinfo {author} {\bibfnamefont {A.}~\bibnamefont {Macridin}}, \bibinfo {author} {\bibfnamefont {M.}~\bibnamefont {Jarrell}}, \bibinfo {author} {\bibfnamefont {T.}~\bibnamefont {Maier}},\ and\ \bibinfo {author} {\bibfnamefont {D.~J.}\ \bibnamefont {Scalapino}},\ }\bibfield  {title} {\bibinfo {title} {High-energy kink in the single-particle spectra of the two-dimensional hubbard model},\ }\href {https://doi.org/10.1103/PhysRevLett.99.237001} {\bibfield  {journal} {\bibinfo  {journal} {Phys. Rev. Lett.}\ }\textbf {\bibinfo {volume} {99}},\ \bibinfo {pages} {237001} (\bibinfo {year} {2007})}\BibitemShut {NoStop}%
\bibitem [{\citenamefont {Moritz}\ \emph {et~al.}(2010)\citenamefont {Moritz}, \citenamefont {Johnston},\ and\ \citenamefont {Devereaux}}]{Moritz2010}%
  \BibitemOpen
  \bibfield  {author} {\bibinfo {author} {\bibfnamefont {B.}~\bibnamefont {Moritz}}, \bibinfo {author} {\bibfnamefont {S.}~\bibnamefont {Johnston}},\ and\ \bibinfo {author} {\bibfnamefont {T.}~\bibnamefont {Devereaux}},\ }\bibfield  {title} {\bibinfo {title} {Insights on the cuprate high energy anomaly observed in arpes},\ }\href {https://doi.org/https://doi.org/10.1016/j.elspec.2010.06.001} {\bibfield  {journal} {\bibinfo  {journal} {Journal of Electron Spectroscopy and Related Phenomena}\ }\textbf {\bibinfo {volume} {181}},\ \bibinfo {pages} {31} (\bibinfo {year} {2010})}\BibitemShut {NoStop}%
\bibitem [{Note6()}]{Note6}%
  \BibitemOpen
  \bibinfo {note} {Also experimentally, the evidence towards $d$-wave superconductivity tightens \cite {Harvey2022,Chow2022}, while --at the beginning-- also mixed $d$-and $s$-wave superconductivity has been suggested \cite {Gu2020b}.}\BibitemShut {Stop}%
\bibitem [{Note7()}]{Note7}%
  \BibitemOpen
  \bibinfo {note} {The extent to which the intrinsic $T_C$ differs in experiment beyond likely differences in the quality of the films is not clear at the moment.}\BibitemShut {Stop}%
\bibitem [{\citenamefont {Inosov}\ \emph {et~al.}(2007)\citenamefont {Inosov}, \citenamefont {Fink}, \citenamefont {Kordyuk}, \citenamefont {Borisenko}, \citenamefont {Zabolotnyy}, \citenamefont {Schuster}, \citenamefont {Knupfer}, \citenamefont {B\"uchner}, \citenamefont {Follath}, \citenamefont {D\"urr}, \citenamefont {Eberhardt}, \citenamefont {Hinkov}, \citenamefont {Keimer},\ and\ \citenamefont {Berger}}]{PhysRevLett.99.237002}%
  \BibitemOpen
  \bibfield  {author} {\bibinfo {author} {\bibfnamefont {D.~S.}\ \bibnamefont {Inosov}}, \bibinfo {author} {\bibfnamefont {J.}~\bibnamefont {Fink}}, \bibinfo {author} {\bibfnamefont {A.~A.}\ \bibnamefont {Kordyuk}}, \bibinfo {author} {\bibfnamefont {S.~V.}\ \bibnamefont {Borisenko}}, \bibinfo {author} {\bibfnamefont {V.~B.}\ \bibnamefont {Zabolotnyy}}, \bibinfo {author} {\bibfnamefont {R.}~\bibnamefont {Schuster}}, \bibinfo {author} {\bibfnamefont {M.}~\bibnamefont {Knupfer}}, \bibinfo {author} {\bibfnamefont {B.}~\bibnamefont {B\"uchner}}, \bibinfo {author} {\bibfnamefont {R.}~\bibnamefont {Follath}}, \bibinfo {author} {\bibfnamefont {H.~A.}\ \bibnamefont {D\"urr}}, \bibinfo {author} {\bibfnamefont {W.}~\bibnamefont {Eberhardt}}, \bibinfo {author} {\bibfnamefont {V.}~\bibnamefont {Hinkov}}, \bibinfo {author} {\bibfnamefont {B.}~\bibnamefont {Keimer}},\ and\ \bibinfo {author} {\bibfnamefont {H.}~\bibnamefont {Berger}},\ }\bibfield  {title} {\bibinfo {title} {Momentum and energy dependence of the anomalous
  high-energy dispersion in the electronic structure of high temperature superconductors},\ }\href {https://doi.org/10.1103/PhysRevLett.99.237002} {\bibfield  {journal} {\bibinfo  {journal} {Phys. Rev. Lett.}\ }\textbf {\bibinfo {volume} {99}},\ \bibinfo {pages} {237002} (\bibinfo {year} {2007})}\BibitemShut {NoStop}%
\bibitem [{\citenamefont {Graf}\ \emph {et~al.}(2007)\citenamefont {Graf}, \citenamefont {Gweon}, \citenamefont {McElroy}, \citenamefont {Zhou}, \citenamefont {Jozwiak}, \citenamefont {Rotenberg}, \citenamefont {Bill}, \citenamefont {Sasagawa}, \citenamefont {Eisaki}, \citenamefont {Uchida}, \citenamefont {Takagi}, \citenamefont {Lee},\ and\ \citenamefont {Lanzara}}]{PhysRevLett.98.067004}%
  \BibitemOpen
  \bibfield  {author} {\bibinfo {author} {\bibfnamefont {J.}~\bibnamefont {Graf}}, \bibinfo {author} {\bibfnamefont {G.-H.}\ \bibnamefont {Gweon}}, \bibinfo {author} {\bibfnamefont {K.}~\bibnamefont {McElroy}}, \bibinfo {author} {\bibfnamefont {S.~Y.}\ \bibnamefont {Zhou}}, \bibinfo {author} {\bibfnamefont {C.}~\bibnamefont {Jozwiak}}, \bibinfo {author} {\bibfnamefont {E.}~\bibnamefont {Rotenberg}}, \bibinfo {author} {\bibfnamefont {A.}~\bibnamefont {Bill}}, \bibinfo {author} {\bibfnamefont {T.}~\bibnamefont {Sasagawa}}, \bibinfo {author} {\bibfnamefont {H.}~\bibnamefont {Eisaki}}, \bibinfo {author} {\bibfnamefont {S.}~\bibnamefont {Uchida}}, \bibinfo {author} {\bibfnamefont {H.}~\bibnamefont {Takagi}}, \bibinfo {author} {\bibfnamefont {D.-H.}\ \bibnamefont {Lee}},\ and\ \bibinfo {author} {\bibfnamefont {A.}~\bibnamefont {Lanzara}},\ }\bibfield  {title} {\bibinfo {title} {Universal high energy anomaly in the angle-resolved photoemission spectra of high temperature superconductors: Possible evidence of spinon
  and holon branches},\ }\href {https://doi.org/10.1103/PhysRevLett.98.067004} {\bibfield  {journal} {\bibinfo  {journal} {Phys. Rev. Lett.}\ }\textbf {\bibinfo {volume} {98}},\ \bibinfo {pages} {067004} (\bibinfo {year} {2007})}\BibitemShut {NoStop}%
\bibitem [{\citenamefont {Xie}\ \emph {et~al.}(2007)\citenamefont {Xie}, \citenamefont {Yang}, \citenamefont {Shen}, \citenamefont {Zhao}, \citenamefont {Ou}, \citenamefont {Wei}, \citenamefont {Gu}, \citenamefont {Arita}, \citenamefont {Qiao}, \citenamefont {Namatame}, \citenamefont {Taniguchi}, \citenamefont {Kaneko}, \citenamefont {Eisaki}, \citenamefont {Tsuei}, \citenamefont {Cheng}, \citenamefont {Vobornik}, \citenamefont {Fujii}, \citenamefont {Rossi}, \citenamefont {Yang},\ and\ \citenamefont {Feng}}]{PhysRevLett.98.147001}%
  \BibitemOpen
  \bibfield  {author} {\bibinfo {author} {\bibfnamefont {B.~P.}\ \bibnamefont {Xie}}, \bibinfo {author} {\bibfnamefont {K.}~\bibnamefont {Yang}}, \bibinfo {author} {\bibfnamefont {D.~W.}\ \bibnamefont {Shen}}, \bibinfo {author} {\bibfnamefont {J.~F.}\ \bibnamefont {Zhao}}, \bibinfo {author} {\bibfnamefont {H.~W.}\ \bibnamefont {Ou}}, \bibinfo {author} {\bibfnamefont {J.}~\bibnamefont {Wei}}, \bibinfo {author} {\bibfnamefont {S.~Y.}\ \bibnamefont {Gu}}, \bibinfo {author} {\bibfnamefont {M.}~\bibnamefont {Arita}}, \bibinfo {author} {\bibfnamefont {S.}~\bibnamefont {Qiao}}, \bibinfo {author} {\bibfnamefont {H.}~\bibnamefont {Namatame}}, \bibinfo {author} {\bibfnamefont {M.}~\bibnamefont {Taniguchi}}, \bibinfo {author} {\bibfnamefont {N.}~\bibnamefont {Kaneko}}, \bibinfo {author} {\bibfnamefont {H.}~\bibnamefont {Eisaki}}, \bibinfo {author} {\bibfnamefont {K.~D.}\ \bibnamefont {Tsuei}}, \bibinfo {author} {\bibfnamefont {C.~M.}\ \bibnamefont {Cheng}}, \bibinfo {author} {\bibfnamefont {I.}~\bibnamefont {Vobornik}},
  \bibinfo {author} {\bibfnamefont {J.}~\bibnamefont {Fujii}}, \bibinfo {author} {\bibfnamefont {G.}~\bibnamefont {Rossi}}, \bibinfo {author} {\bibfnamefont {Z.~Q.}\ \bibnamefont {Yang}},\ and\ \bibinfo {author} {\bibfnamefont {D.~L.}\ \bibnamefont {Feng}},\ }\bibfield  {title} {\bibinfo {title} {High-energy scale revival and giant kink in the dispersion of a cuprate superconductor},\ }\href {https://doi.org/10.1103/PhysRevLett.98.147001} {\bibfield  {journal} {\bibinfo  {journal} {Phys. Rev. Lett.}\ }\textbf {\bibinfo {volume} {98}},\ \bibinfo {pages} {147001} (\bibinfo {year} {2007})}\BibitemShut {NoStop}%
\bibitem [{\citenamefont {Meevasana}\ \emph {et~al.}(2007)\citenamefont {Meevasana}, \citenamefont {Zhou}, \citenamefont {Sahrakorpi}, \citenamefont {Lee}, \citenamefont {Yang}, \citenamefont {Tanaka}, \citenamefont {Mannella}, \citenamefont {Yoshida}, \citenamefont {Lu}, \citenamefont {Chen}, \citenamefont {He}, \citenamefont {Lin}, \citenamefont {Komiya}, \citenamefont {Ando}, \citenamefont {Zhou}, \citenamefont {Ti}, \citenamefont {Xiong}, \citenamefont {Zhao}, \citenamefont {Sasagawa}, \citenamefont {Kakeshita}, \citenamefont {Fujita}, \citenamefont {Uchida}, \citenamefont {Eisaki}, \citenamefont {Fujimori}, \citenamefont {Hussain}, \citenamefont {Markiewicz}, \citenamefont {Bansil}, \citenamefont {Nagaosa}, \citenamefont {Zaanen}, \citenamefont {Devereaux},\ and\ \citenamefont {Shen}}]{PhysRevB.75.174506}%
  \BibitemOpen
  \bibfield  {author} {\bibinfo {author} {\bibfnamefont {W.}~\bibnamefont {Meevasana}}, \bibinfo {author} {\bibfnamefont {X.~J.}\ \bibnamefont {Zhou}}, \bibinfo {author} {\bibfnamefont {S.}~\bibnamefont {Sahrakorpi}}, \bibinfo {author} {\bibfnamefont {W.~S.}\ \bibnamefont {Lee}}, \bibinfo {author} {\bibfnamefont {W.~L.}\ \bibnamefont {Yang}}, \bibinfo {author} {\bibfnamefont {K.}~\bibnamefont {Tanaka}}, \bibinfo {author} {\bibfnamefont {N.}~\bibnamefont {Mannella}}, \bibinfo {author} {\bibfnamefont {T.}~\bibnamefont {Yoshida}}, \bibinfo {author} {\bibfnamefont {D.~H.}\ \bibnamefont {Lu}}, \bibinfo {author} {\bibfnamefont {Y.~L.}\ \bibnamefont {Chen}}, \bibinfo {author} {\bibfnamefont {R.~H.}\ \bibnamefont {He}}, \bibinfo {author} {\bibfnamefont {H.}~\bibnamefont {Lin}}, \bibinfo {author} {\bibfnamefont {S.}~\bibnamefont {Komiya}}, \bibinfo {author} {\bibfnamefont {Y.}~\bibnamefont {Ando}}, \bibinfo {author} {\bibfnamefont {F.}~\bibnamefont {Zhou}}, \bibinfo {author} {\bibfnamefont {W.~X.}\ \bibnamefont {Ti}},
  \bibinfo {author} {\bibfnamefont {J.~W.}\ \bibnamefont {Xiong}}, \bibinfo {author} {\bibfnamefont {Z.~X.}\ \bibnamefont {Zhao}}, \bibinfo {author} {\bibfnamefont {T.}~\bibnamefont {Sasagawa}}, \bibinfo {author} {\bibfnamefont {T.}~\bibnamefont {Kakeshita}}, \bibinfo {author} {\bibfnamefont {K.}~\bibnamefont {Fujita}}, \bibinfo {author} {\bibfnamefont {S.}~\bibnamefont {Uchida}}, \bibinfo {author} {\bibfnamefont {H.}~\bibnamefont {Eisaki}}, \bibinfo {author} {\bibfnamefont {A.}~\bibnamefont {Fujimori}}, \bibinfo {author} {\bibfnamefont {Z.}~\bibnamefont {Hussain}}, \bibinfo {author} {\bibfnamefont {R.~S.}\ \bibnamefont {Markiewicz}}, \bibinfo {author} {\bibfnamefont {A.}~\bibnamefont {Bansil}}, \bibinfo {author} {\bibfnamefont {N.}~\bibnamefont {Nagaosa}}, \bibinfo {author} {\bibfnamefont {J.}~\bibnamefont {Zaanen}}, \bibinfo {author} {\bibfnamefont {T.~P.}\ \bibnamefont {Devereaux}},\ and\ \bibinfo {author} {\bibfnamefont {Z.-X.}\ \bibnamefont {Shen}},\ }\bibfield  {title} {\bibinfo {title} {Hierarchy of
  multiple many-body interaction scales in high-temperature superconductors},\ }\href {https://doi.org/10.1103/PhysRevB.75.174506} {\bibfield  {journal} {\bibinfo  {journal} {Phys. Rev. B}\ }\textbf {\bibinfo {volume} {75}},\ \bibinfo {pages} {174506} (\bibinfo {year} {2007})}\BibitemShut {NoStop}%
\bibitem [{\citenamefont {Ronning}\ \emph {et~al.}(2005)\citenamefont {Ronning}, \citenamefont {Shen}, \citenamefont {Armitage}, \citenamefont {Damascelli}, \citenamefont {Lu}, \citenamefont {Shen}, \citenamefont {Miller},\ and\ \citenamefont {Kim}}]{PhysRevB.71.094518}%
  \BibitemOpen
  \bibfield  {author} {\bibinfo {author} {\bibfnamefont {F.}~\bibnamefont {Ronning}}, \bibinfo {author} {\bibfnamefont {K.~M.}\ \bibnamefont {Shen}}, \bibinfo {author} {\bibfnamefont {N.~P.}\ \bibnamefont {Armitage}}, \bibinfo {author} {\bibfnamefont {A.}~\bibnamefont {Damascelli}}, \bibinfo {author} {\bibfnamefont {D.~H.}\ \bibnamefont {Lu}}, \bibinfo {author} {\bibfnamefont {Z.-X.}\ \bibnamefont {Shen}}, \bibinfo {author} {\bibfnamefont {L.~L.}\ \bibnamefont {Miller}},\ and\ \bibinfo {author} {\bibfnamefont {C.}~\bibnamefont {Kim}},\ }\bibfield  {title} {\bibinfo {title} {Anomalous high-energy dispersion in angle-resolved photoemission spectra from the insulating cuprate ${\mathrm{ca}}_{2}{\mathrm{cuo}}_{2}{\mathrm{cl}}_{2}$},\ }\href {https://doi.org/10.1103/PhysRevB.71.094518} {\bibfield  {journal} {\bibinfo  {journal} {Phys. Rev. B}\ }\textbf {\bibinfo {volume} {71}},\ \bibinfo {pages} {094518} (\bibinfo {year} {2005})}\BibitemShut {NoStop}%
\bibitem [{\citenamefont {Chen}\ \emph {et~al.}(2022)\citenamefont {Chen}, \citenamefont {Osada}, \citenamefont {Li}, \citenamefont {Been}, \citenamefont {Chen}, \citenamefont {Hashimoto}, \citenamefont {Lu}, \citenamefont {Mo}, \citenamefont {Lee}, \citenamefont {Wang}, \citenamefont {Rodolakis}, \citenamefont {McChesney}, \citenamefont {Jia}, \citenamefont {Moritz}, \citenamefont {Devereaux}, \citenamefont {Hwang},\ and\ \citenamefont {Shen}}]{Chen2022}%
  \BibitemOpen
  \bibfield  {author} {\bibinfo {author} {\bibfnamefont {Z.}~\bibnamefont {Chen}}, \bibinfo {author} {\bibfnamefont {M.}~\bibnamefont {Osada}}, \bibinfo {author} {\bibfnamefont {D.}~\bibnamefont {Li}}, \bibinfo {author} {\bibfnamefont {E.~M.}\ \bibnamefont {Been}}, \bibinfo {author} {\bibfnamefont {S.-D.}\ \bibnamefont {Chen}}, \bibinfo {author} {\bibfnamefont {M.}~\bibnamefont {Hashimoto}}, \bibinfo {author} {\bibfnamefont {D.}~\bibnamefont {Lu}}, \bibinfo {author} {\bibfnamefont {S.-K.}\ \bibnamefont {Mo}}, \bibinfo {author} {\bibfnamefont {K.}~\bibnamefont {Lee}}, \bibinfo {author} {\bibfnamefont {B.~Y.}\ \bibnamefont {Wang}}, \bibinfo {author} {\bibfnamefont {F.}~\bibnamefont {Rodolakis}}, \bibinfo {author} {\bibfnamefont {J.~L.}\ \bibnamefont {McChesney}}, \bibinfo {author} {\bibfnamefont {C.}~\bibnamefont {Jia}}, \bibinfo {author} {\bibfnamefont {B.}~\bibnamefont {Moritz}}, \bibinfo {author} {\bibfnamefont {T.~P.}\ \bibnamefont {Devereaux}}, \bibinfo {author} {\bibfnamefont {H.~Y.}\ \bibnamefont
  {Hwang}},\ and\ \bibinfo {author} {\bibfnamefont {Z.-X.}\ \bibnamefont {Shen}},\ }\bibfield  {title} {\bibinfo {title} {Electronic structure of superconducting nickelates probed by resonant photoemission spectroscopy},\ }\href {https://doi.org/https://doi.org/10.1016/j.matt.2022.01.020} {\bibfield  {journal} {\bibinfo  {journal} {Matter}\ }\textbf {\bibinfo {volume} {5}},\ \bibinfo {pages} {1806} (\bibinfo {year} {2022})}\BibitemShut {NoStop}%
\bibitem [{\citenamefont {Harvey}\ \emph {et~al.}(2022)\citenamefont {Harvey}, \citenamefont {Wang}, \citenamefont {Fowlie}, \citenamefont {Osada}, \citenamefont {Lee}, \citenamefont {Lee}, \citenamefont {Li},\ and\ \citenamefont {Hwang}}]{Harvey2022}%
  \BibitemOpen
  \bibfield  {author} {\bibinfo {author} {\bibfnamefont {S.~P.}\ \bibnamefont {Harvey}}, \bibinfo {author} {\bibfnamefont {B.~Y.}\ \bibnamefont {Wang}}, \bibinfo {author} {\bibfnamefont {J.}~\bibnamefont {Fowlie}}, \bibinfo {author} {\bibfnamefont {M.}~\bibnamefont {Osada}}, \bibinfo {author} {\bibfnamefont {K.}~\bibnamefont {Lee}}, \bibinfo {author} {\bibfnamefont {Y.}~\bibnamefont {Lee}}, \bibinfo {author} {\bibfnamefont {D.}~\bibnamefont {Li}},\ and\ \bibinfo {author} {\bibfnamefont {H.~Y.}\ \bibnamefont {Hwang}},\ }\bibfield  {title} {\bibinfo {title} {Evidence for nodal superconductivity in infinite-layer nickelates},\ }\href {https://arxiv.org/abs/2201.12971} {\bibfield  {journal} {\bibinfo  {journal} {arXiv:2201.12971}\ } (\bibinfo {year} {2022})}\BibitemShut {NoStop}%
\bibitem [{\citenamefont {Chow}\ \emph {et~al.}(2022)\citenamefont {Chow}, \citenamefont {Sudheesh}, \citenamefont {Nandi}, \citenamefont {Zeng}, \citenamefont {Zhang}, \citenamefont {Du}, \citenamefont {Lim}, \citenamefont {Chia},\ and\ \citenamefont {Ariando}}]{Chow2022}%
  \BibitemOpen
  \bibfield  {author} {\bibinfo {author} {\bibfnamefont {L.~E.}\ \bibnamefont {Chow}}, \bibinfo {author} {\bibfnamefont {S.~K.}\ \bibnamefont {Sudheesh}}, \bibinfo {author} {\bibfnamefont {P.}~\bibnamefont {Nandi}}, \bibinfo {author} {\bibfnamefont {S.~W.}\ \bibnamefont {Zeng}}, \bibinfo {author} {\bibfnamefont {Z.~T.}\ \bibnamefont {Zhang}}, \bibinfo {author} {\bibfnamefont {X.~M.}\ \bibnamefont {Du}}, \bibinfo {author} {\bibfnamefont {Z.~S.}\ \bibnamefont {Lim}}, \bibinfo {author} {\bibfnamefont {E.~E.~M.}\ \bibnamefont {Chia}},\ and\ \bibinfo {author} {\bibfnamefont {A.}~\bibnamefont {Ariando}},\ }\bibfield  {title} {\bibinfo {title} {Pairing symmetry in infinite-layer nickelate superconductor},\ }\href {https://arxiv.org/abs/2201.10038} {\bibfield  {journal} {\bibinfo  {journal} {arXiv:2201.10038}\ } (\bibinfo {year} {2022})}\BibitemShut {NoStop}%
\bibitem [{\citenamefont {Gu}\ \emph {et~al.}(2020)\citenamefont {Gu}, \citenamefont {Li}, \citenamefont {Wan}, \citenamefont {Li}, \citenamefont {Guo}, \citenamefont {Yang}, \citenamefont {Li}, \citenamefont {Zhu}, \citenamefont {Pan}, \citenamefont {Nie},\ and\ \citenamefont {Wen}}]{Gu2020b}%
  \BibitemOpen
  \bibfield  {author} {\bibinfo {author} {\bibfnamefont {Q.}~\bibnamefont {Gu}}, \bibinfo {author} {\bibfnamefont {Y.}~\bibnamefont {Li}}, \bibinfo {author} {\bibfnamefont {S.}~\bibnamefont {Wan}}, \bibinfo {author} {\bibfnamefont {H.}~\bibnamefont {Li}}, \bibinfo {author} {\bibfnamefont {W.}~\bibnamefont {Guo}}, \bibinfo {author} {\bibfnamefont {H.}~\bibnamefont {Yang}}, \bibinfo {author} {\bibfnamefont {Q.}~\bibnamefont {Li}}, \bibinfo {author} {\bibfnamefont {X.}~\bibnamefont {Zhu}}, \bibinfo {author} {\bibfnamefont {X.}~\bibnamefont {Pan}}, \bibinfo {author} {\bibfnamefont {Y.}~\bibnamefont {Nie}},\ and\ \bibinfo {author} {\bibfnamefont {H.-H.}\ \bibnamefont {Wen}},\ }\bibfield  {title} {\bibinfo {title} {{Single particle tunneling spectrum of superconducting {Nd$_{1-x}$Sr$_x$NiO$_2$} thin films}},\ }\href {https://doi.org/10.1038/s41467-020-19908-1} {\bibfield  {journal} {\bibinfo  {journal} {Nature Comm.}\ }\textbf {\bibinfo {volume} {11}},\ \bibinfo {pages} {6027} (\bibinfo {year} {2020})}\BibitemShut
  {NoStop}%
\end{thebibliography}

%

\end{document}